\documentclass{article}%
\usepackage{amssymb}
\usepackage{amsfonts}
\usepackage{amsmath}
\usepackage[numbers,sort&compress]{natbib}
\usepackage{graphicx}%
\providecommand{\U}[1]{\protect\rule{.1in}{.1in}}
\newtheorem{theorem}{Theorem}

\newtheorem{example}{Example}

\newtheorem{lemma}{Lemma}

\newtheorem{problem}{Problem}

\newtheorem{remark}{Remark}

\numberwithin{equation}{section}
\begin{document}

\title{Systematic construction of non-autonomous Hamiltonian equations of
Painlev\'{e}-type. I. Frobenius integrability }
\author{Maciej B\l aszak\\Faculty of Physics, Department of Mathematical Physics and Computer Modelling,\\A. Mickiewicz University\\Uniwersytetu Pozna\'nskiego 2, 61-614 Pozna\'{n}, Poland\\\texttt{blaszakm@amu.edu.pl}
\and Krzysztof Marciniak\\Department of Science and Technology \\Campus Norrk\"{o}ping, Link\"{o}ping University\\601-74 Norrk\"{o}ping, Sweden\\\texttt{krzma@itn.liu.se}
\and Ziemowit Doma\'nski\\Institute of Mathematics, Pozna{\'{n}} University of Technology\\Piotrowo 3A, 60-965 Pozna{\'{n}}, Poland\\\texttt{ziemowit.domanski@put.poznan.pl}}
\maketitle

\begin{abstract}
This article is the first one in a suite of three articles exploring
connections between dynamical systems of St\"{a}ckel-type and of Painlev\'{e}-
type. In this article we present a deformation of autonomous St\"{a}ckel-type
systems to non-autonomous Frobenius integrable systems. First, we consider
quasi-St\"{a}ckel systems with quadratic in momenta Hamiltonians containing
separable potentials with time dependent coefficients and then we present a
procedure of deforming these equations to non-autonomous Frobenius integrable
systems. Then, we present a procedure of deforming quasi-St\"{a}ckel systems
with so called magnetic separable potentials to non-autonomous Frobenius
integrable systems. We also provide a complete list of all $2$- and
$3\,$-dimensional Frobenius integrable systems, both with ordinary and with
magnetic potentials, that originate in our construction. Further, we prove the
equivalence between both classes of systems. Finally we show how Painlev\'{e}
equations $P_{I}-P_{IV}$ can be derived from our scheme.

\end{abstract}



\section{ Introduction}

Among all second order nonlinear ordinary differential equations (ODE's) there
are two distinguished classes, playing important roles in modern mathematics
and physics. The first class is represented by nonlinear equations of
St\"{a}ckel-type, with an autonomous Hamiltonian representation (see
\cite{Book} and references therein). The second class is represented by
nonlinear ordinary differential equations of Painlev\'{e}-type, with a
non-autonomous Hamiltonian representation (see \cite{Iwasaki} and references
therein). In both cases Hamiltonian functions are quadratic in momenta. This
paper is a first paper in a suite of three papers exploring the so far unknown
--- to the best knowledge of the authors --- connections between both types of systems.

Let us briefly characterize the systems under consideration. On a
$2n$-dimensional Poisson manifold $(\mathcal{M},\pi)$, where $\pi$ is a
non-degenerated Poisson bi-vector (so that $M$ is in fact a symplectic
manifold with the associated Poisson bivector $\pi$), consider a set of $n$
autonomous evolution equations (autonomous dynamical systems) of the form%
\begin{equation}
\frac{d\xi}{dt_{r}}=X_{r}(\xi)\equiv\pi dh_{r}(\xi),\quad r=1,\ldots,n,
\label{0}%
\end{equation}
where $\xi\in\mathcal{M}$ denotes points on $\mathcal{M},$ $h_{r}$ are
Hamiltonian functions (smooth real-valued functions on $\mathcal{M}$) and
$X_{r}=\pi dh_{r}$ are the related Hamiltonian vector fields on $\mathcal{M}$.
The set of $n$ equations (\ref{0}) constitutes an autonomous
\emph{St\"{a}ckel-type system} if the following two conditions are satisfied:

\begin{enumerate}
\item All $n$ Hamiltonian functions Poisson-commute%
\begin{equation}
\left\{  h_{r},h_{s}\right\}  :=\pi(dh_{r},dh_{s})=0,\quad r,s=1,\ldots,n,
\label{kom0}%
\end{equation}
so the system is Liouville integrable. In consequence all the vector fields
$X_{r\text{ }}$commute as well
\[
\left[  X_{r},X_{s}\right]  =0,\quad r,s=1,\ldots,n
\]
and hence the system (\ref{0}) has a common, unique (at least local) solution
$\xi(t_{1},\ldots,t_{n},\xi_{0})$ through each point $\xi_{0}\in\mathcal{M},$
depending in general on all the evolution parameters $t_{s}$.

\item The autonomous equations (\ref{0}) are represented by (i.e. are
differential consequences of) the Lax equations
\begin{equation}
\frac{d}{dt_{k}}L(x;\xi)=[U_{k}(x;\xi),L(x;\xi)],\quad k=1,\ldots,n, \label{8}%
\end{equation}
where
\[
\frac{d}{dt_{k}}=\{\,\cdot\,,h_{k}\}
\]
is the evolutionary derivative along the flow $k$ in (\ref{0}) and with
$L(x;\xi)$ and $U_{k}(x;\xi)$ being matrices belonging to some Lie algebra,
depending rationally on the parameter $x$ called a spectral parameter.
\end{enumerate}

Equations (\ref{8}) are called the isospectral deformation equations, as the
eigenvalues of the matrix $L$ are independent of all times $t_{k}$,
$k=1,\dotsc,n.$ The Lax representation (\ref{8}) allows to find an explicit
form of transformation to the so called \emph{separation coordinates }and in
consequence\emph{\ }to solve all the evolution equations (\ref{0}) in
quadratures (see \cite{Ad,Sk,Ei} and references therein). One can also invert
the whole procedure and start the construction of St\"{a}ckel-type system from
particular separation relations, as it is done in this paper (see Section 2 as
well as \cite{Book,BD2019,BM}).

Now, on the same symplectic/Poisson manifold $(\mathcal{M},\pi)$, consider a
set of $n$ non-autonomous evolution equations (non-autonomous dynamical
systems) of the form%
\begin{equation}
\frac{d\xi}{dt_{r}}=Y_{r}(\xi,t)=\pi dH_{r}(\xi,t),\quad r=1,\ldots,n,
\label{Pff}%
\end{equation}
where $t=(t_{1},\ldots,t_{n})$. The set of $n$ equations (\ref{Pff})
constitutes a non-autonomous \emph{Painlev\'{e}-type system} if the following
two conditions are satisfied:

\begin{enumerate}
\item Hamiltonian functions $H_{r}$ fulfill the Frobenius integrability
condition%
\begin{equation}
\frac{\partial H_{r}}{\partial t_{s}}-\frac{\partial H_{s}}{\partial t_{r}%
}+\{H_{r},H_{s}\}=f_{rs}(t_{1},\ldots,t_{n}),\quad r,s=1,\dots,n, \label{fcg}%
\end{equation}
(cf. (\ref{kom0})) where $f_{rs}$ are some functions not depending on the
phase-space variables $\xi$, only on the parameters $t_{j}$ (note that in
general the left hand side could also depend on $\xi$). By an appropriate
shifting $H_{r}\rightarrow H_{r}+c_{r}(t)$ we can always make all the
functions $f_{rs}$ vanish. Due to (\ref{fcg}) the non-autonomous Hamiltonian
vector fields $Y_{k}(\xi,t)$ satisfy
\begin{equation}
\frac{\partial Y_{r}}{\partial t_{s}}-\frac{\partial Y_{s}}{\partial t_{r}%
}+\left[  Y_{s},Y_{r}\right]  =0,\quad r,s=1,\dots,n, \label{fcv}%
\end{equation}
as $\pi d\{H_{r},H_{s}\}=-\left[  Y_{r},Y_{s}\right]  $. Therefore, the set of
non-autonomous evolution equations (\ref{Pff}) has again common (at least
local) solutions $\xi(t_{1},\ldots,t_{n},\xi_{0})$ through each point $\xi
_{0}$ of $\mathcal{M}$ . For the notion of Frobenius integrability, see for
example \cite{Fecko,Iwasaki,Lundell}.

\item The non-autonomous Hamiltonian equations (\ref{Pff}) are represented by
(i.e. are differential consequences of) the isomonodromic Lax representation
\begin{equation}
\frac{d}{dt_{k}}L(x;\xi,t)=[U_{k}(x;\xi,t),L(x;\xi,t)]+\frac{\partial
U_{k}(x;\xi,t)}{\partial x},\quad k=1,\ldots,n \label{iso}%
\end{equation}
where matrices $L$ and $U_{k}$ have rational singularities in $x$, and where
now
\[
\frac{d}{dt_{k}}=\frac{\partial}{\partial t_{k}}+\{\,\cdot\,,h_{k}\}
\]
is the evolutionary derivative along the flow $k.$
\end{enumerate}

Note that the isomonodromic Lax representation (\ref{iso}) is only a necessary
condition for the Painlev\'{e} property \cite{D}; this is why the system
(\ref{Pff}) with the property (\ref{fcg}) and the representation (\ref{iso})
is called Painlev\'{e}-type.

Nowadays we have a thorough knowledge of separable St\"{a}ckel-type systems
(\ref{0}). We know for example how to construct them from appropriate
separation relations, together with their Lax representations for arbitrary
$n$ (see \cite{Book,BD2019} and references therein). Also, a significant
progress in constructing new multi-component Painlev\'{e}-type equations took
place since the modern theory of nonlinear integrable PDE's has been born (the
so-called \emph{soliton theory}), starting from the seminal papers
\cite{Jim1,Jim2,Jim3}. It was found that Painlev\'{e} equations are strongly
connected with the soliton systems and they share many of their properties
(see \cite{S1,S2,S3,S4,S5,t} and references therein). In particular,
Painlev\'{e} equations were constructed under various similarity reductions of
soliton PDE's hierarchies but mostly only a single equation from the family
(\ref{Pff}), coming from $H_{1}$, was derived. However, there is virtually no
papers/no results on \emph{connections/relations} between these two types of systems.

This paper is the first in a suite of papers that exploit relations between
both type of systems. The main objective of this paper is investigations of
systems belonging to first of the above definitions, i.e. a systematic
construction of non-autonomous Frobenius integrable systems by appropriate
deformations of St\"{a}ckel-type systems. This method has been introduced in
\cite{arxiv} (see also \cite{M2019} for a simple illustration of this method).
Their Lax representations will be derived in the next paper of the suite.

This paper is based on some previous results which we briefly mention here. In
\cite{mb} we introduced and investigated the so called geodesic
quasi-St\"{a}ckel Hamiltonians. These Hamiltonians constitute a non-commuting
finite-dimensional Lie algebra with respect to the Poisson bracket and thus
evolution equations they generate are not Frobenius integrable. Then, in
\cite{arxiv} we proved how to deform this algebra to a set of non-autonomous
Hamiltonians such that related systems are integrable in the Frobenius sense
and such that both sets of associated Hamiltonian vector fields, these before
the deformation and these after the deformation, span the same distribution on
$\mathcal{M}$.

This paper is organized as follows. In Section \ref{sec2} we briefly remind
the definition and main properties of Liouville integrable systems of
St\"{a}ckel-type. In Section \ref{bla} we describe, following the results in
\cite{mb,arxiv}, the construction of autonomous geodesic quasi-St\"{a}ckel
systems as well as their non-autonomous deformations preserving Frobenius
integrability. Sections from \ref{sec4} and onwards contain almost solely new
results. In Section \ref{sec4} we introduce quasi-St\"{a}ckel systems with
admissible separable potentials and present a procedure of their deformation
into non-autonomous Frobenius integrable systems. In Section \ref{sec5} we
present a classification of two- and three-dimensional Frobenius integrable
non-autonomous Hamiltonian ODE's with potentials obtained in this way. In
Section \ref{sec6} we introduce quasi-St\"{a}ckel systems with admissible
separable potentials that we call \emph{magnetic} potentials (for the reasons
explained in Section \ref{sec6}) and present a procedure of their deformation
into non-autonomous Frobenius integrable equations. In Section \ref{sec7} we
present a classification of two- and three-dimensional Frobenius integrable
non-autonomous Hamiltonian systems with magnetic potentials. In Section
\ref{sec8} we construct a multitime-dependent canonical transformation between
non-autonomous systems with magnetic potentials and those with ordinary
potentials. The final section of this paper is devoted to constructing
Painlev\'{e} $P_{I}-P_{IV}$ equations in the framework of the presented formalism.

\section{Geodesic St\"{a}ckel systems\label{sec2}}

Fix $n\in\mathbb{N}$. Consider a $2n$-dimensional Poisson manifold
$(\mathcal{M},\pi)$ with a non-degenerate Poisson bivector $\pi$ (so that
$\mathcal{M}$ is in fact a symplectic manifold with the associated Poisson
bivector $\pi$) and a particular set $(\lambda,\mu)=(\lambda_{1}%
,\ldots,\lambda_{n},\allowbreak\mu_{1},\ldots,\mu_{n})$ of \emph{almost
global} Darboux (canonical) coordinates on $\mathcal{M}$ (we assume that they
exist), so that $\{\mu_{i},\lambda_{j}\}_{\pi}=\delta_{ij}$, $\ \{\lambda
_{i},\lambda_{j}\}_{\pi}=\{\mu_{i},\mu_{j}\}_{\pi}=0$, $i,j=1,\dots,n$. By
\emph{almost global }we mean defined everywhere except on a zero-measure set
given by some polynomial equations (zeros of functions $\Delta_{i}$ on
$\mathcal{M}$, defined below).

Now, for a fixed $m\in\mathbb{Z}$, consider the following algebraic separation
curve in the $(x,y)$-plane
\begin{equation}
\sum_{r=1}^{n}E_{r}x^{n-r}=\frac{1}{2}x^{m}y^{2}. \label{1}%
\end{equation}
By taking $n$ copies of (\ref{1}) at points $(x,y)=(\lambda_{i},\mu_{i})$,
$i=1,\dotsc,n$, we obtain a system of $n$ linear equations (separation
relations) for $E_{r}$%
\begin{equation}
\sum_{r=1}^{n}E_{r}\lambda_{i}^{n-r}=\frac{1}{2}\lambda_{i}^{m}\mu_{i}%
^{2},\ \ \ i=1,...,n. \label{1a}%
\end{equation}
Solving (\ref{1a}) with respect to $E_{r}$ requires inverting the St\"{a}ckel
matrix $S_{ir}=\lambda_{i}^{n-r}$ that in this case has a form of the
Vandermonde matrix.

\begin{lemma}
\label{VDM}\bigskip If $S$ is the $n\times n$ Vandermonde matrix given by
$S_{ij}=\lambda_{i}^{n-j}$ then%
\[
\left[  S^{-1}\right]  _{ij}=-\frac{1}{\Delta_{j}}\frac{\partial\rho_{i}%
}{\partial\lambda_{j}}%
\]
where%
\[
\rho_{i}=(-1)^{i}s_{i}(\lambda),\ \ \ \ \Delta_{j}=\prod\limits_{k\neq
j}(\lambda_{j}-\lambda_{k})
\]
and where $s_{r}\left(  \lambda\right)  $ are elementary symmetric polynomials.
\end{lemma}

This lemma can be proved by direct calculation. By the above lemma, solving
(\ref{1}) with respect to $E_{r}$ yields $n$ functions (Hamiltonians) $E_{r}$
on $(\mathcal{M},\pi)$
\begin{equation}
E_{r}=-\frac{1}{2}\sum_{i=1}^{n}\frac{\partial\rho_{r}}{\partial\lambda_{i}%
}\frac{\lambda_{i}^{m}\mu_{i}^{2}}{\Delta_{i}}\equiv\frac{1}{2}\mu^{T}%
K_{r}G\mu,\quad r=1,\ldots,n \label{4}%
\end{equation}
where%
\[
G=\text{diag}\left(  \frac{\lambda_{1}^{m}}{\Delta_{1}},\ldots,\frac
{\lambda_{n}^{m}}{\Delta_{n}}\right)
\]
and
\[
K_{r}=-\text{diag}\left(  \frac{\partial\rho_{r}}{\partial\lambda_{1}}%
,\cdots,\frac{\partial\rho_{r}}{\partial\lambda_{n}}\right)  ,\quad
r=1,\ldots,n.
\]
The Hamiltonians $E_{r}$ depend on $m$ through the $m$-dependent matrix $G$
that can be interpreted as a contravariant metric tensor on an $n$-dimensional
manifold $Q$ such that $\mathcal{M}=T^{\ast}Q$ (which means that $g=G^{-1}$ is
considered as a usual, covariant metric on $\mathcal{M}$). It can be shown
that the metric $g=G^{-1}$ is flat for $m=0,\dotsc,n$ and of constant
curvature for $m=n+1$. Matrices $K_{r}$ --- note that they do not depend on
$m$ --- can be shown to be $(1,1)$-Killing tensors for the metric $G$ for
arbitrary $m\in\mathbb{Z}$. The first Hamiltonian $E_{1}$ can then be
interpreted as the Hamiltonian of a free particle in the pseudo-Riemannian
configuration space $(Q,g=G^{-1})$ \cite{blasz2005,blasz2007,Book}. Note that
since $G$ is a $(2,0)$-tensor and $K_{r}$ is a $(1,1)$-tensor on $Q$ then
$K_{r}G$ is again a $(2,0)$-tensor on $Q$ while $\mu^{T}K_{r}G\mu$ is a
quadratic form on $\mathcal{M}=T^{\ast}Q$ i.e. a function on $\mathcal{M}$.

By their very construction from separation relations, the Hamiltonians $E_{r}$
(for any fixed $m$) Poisson-commute for all $r,s=1,\ldots,n$
\[
\{E_{r},E_{s}\}_{\pi}=\pi(dE_{r},dE_{s})=0
\]
(see for example \cite{Book}, p. 140), so that $[X_{r},X_{s}]=0$ where $X_{r}$
denote $n$ related Hamiltonian vector fields $X_{r}=\pi dE_{r},$
$r=1,\dotsc,n$. Thus, for each fixed choice of $m$, the $n$ Hamiltonians
$E_{r}$, $r=1,\ldots,n$, constitute a geodesic St\"{a}ckel system from the so
called Benenti class \cite{Ben1,Ben2,blasz2005}. The Darboux coordinates
$(\lambda,\mu)$ are separation coordinates for $E_{r}$.

\section{Frobenius integrable deformations of geodesic quasi-St\"{a}ckel
systems\label{bla}}

In this section we briefly remind, following \cite{mb} and \cite{arxiv}, the
construction of autonomous geodesic quasi-St\"{a}ckel systems and their
non-autonomous Frobenius integrable deformations.

Fix $m\in\{0,\dots,n+1\}$ and consider the following system of
quasi-separation relations (cf.\ also \cite{marikhin 2014} and compare with
\cite{marikhin 2005}) on $\mathcal{M}$
\begin{equation}
\sum_{r=1}^{n}\lambda_{i}^{n-r}\mathcal{E}_{r}=\frac{1}{2}\lambda_{i}^{m}%
\mu_{i}^{2}+\sum_{k=1}^{n}v_{ik}(\lambda)\mu_{k},\quad i=1,\ldots,n,
\label{sep}%
\end{equation}
where $v_{ik}$ are some, unspecified so far, functions of all $\lambda_{i}.$
Solving (\ref{sep}) with respect to $\mathcal{E}_{r}$ yields, for each choice
of $m\in\{0,\dots,n+1\}$, $n$ Hamiltonians on $\mathcal{M}$:
\begin{equation}
\mathcal{E}_{r}=\ E_{r}+W_{r}=\frac{1}{2}\mu^{T}GK_{r}\mu+\mu^{T}J_{r},\quad
r=1,\dotsc,n. \label{qsep}%
\end{equation}
with $E_{r}$ exactly as in (\ref{4}). The additional terms $W_{r}=\mu^{T}%
J_{r}=\sum\limits_{i=1}^{n}J_{r}^{i}\mu_{i}$ (we will refer to them as
\emph{quasi-St\"{a}ckel terms}) are nonseparable linear in momenta functions
on $\mathcal{M}=T^{\ast}Q$ induced by the vector fields $J_{r}=J_{r}^{i}%
\frac{\partial}{\partial\lambda_{i}}$ on $Q$.

Let us impose two conditions on Hamiltonians (\ref{qsep}):

\begin{enumerate}
\item $W_{1}=0$ and $\{E_{1},W_{r}\}=0$, $r=2,\dotsc,n$, which means that the
vector fields $J_{r}$ are Killing vector fields for $G.$

\item The Hamiltonians $\mathcal{E}_{r}$ constitute a Lie algebra with respect
to Poisson bracket.
\end{enumerate}

The first condition is met only for $m\in\{0,\dots,n+1\},$ as then the metric
tensor $G$ has a sufficient number of Killing vectors. Further, both
conditions are met if (a sufficient condition) the functions $v_{ik}$ are
chosen as%

\begin{equation}
\sum_{k=1}^{n}v_{ik}(\lambda)\mu_{k}=%
\begin{cases}
\displaystyle-\sum_{k\neq i}\frac{\mu_{i}-\mu_{k}}{\lambda_{i}-\lambda_{k}}, &
\ \text{for }m=0,\\[5mm]%
\displaystyle-\lambda_{i}^{m-1}\sum_{k\neq i}\frac{\lambda_{i}\mu_{i}%
-\lambda_{k}\mu_{k}}{\lambda_{i}-\lambda_{k}}+(m-1)\lambda_{i}^{m-1}\mu_{i}, &
\ \text{for }m=1,\ldots,n,\\[5mm]%
\displaystyle-\lambda_{i}^{n-1}\sum_{k\neq i}\frac{\lambda_{i}^{2}\mu
_{i}-\lambda_{k}^{2}\mu_{k}}{\lambda_{i}-\lambda_{k}}+(n-1)\lambda_{i}^{n}%
\mu_{i}, & \ \text{for }m=n+1.
\end{cases}
\label{ni}%
\end{equation}
The formulas (\ref{ni}) constitute a definition of the set of functions
$v_{ik}$ for various $m$ (thus $v_{ik}$ depend on $m\,$). With this choice of
$v_{ik}$ the components of Killing vector fields $J_{r}$ are given explicitly
by \cite{mb}
\begin{equation}
J_{r}^{i}=-\sum\limits_{k=1}^{r-1}\,k\,\rho_{r-k-1}\frac{\lambda_{i}^{m+k-1}%
}{\Delta_{i}},\quad r\in I_{1}^{m} \label{j1}%
\end{equation}
and
\begin{equation}
J_{r}^{i}=-\sum\limits_{k=1}^{n-r+1}\,k\,\rho_{r+k-1}\frac{\lambda_{i}%
^{m-k-1}}{\Delta_{i}},\quad r\in I_{2}^{m}, \label{j2}%
\end{equation}
where for each $m\in\{0,\ldots,n+1\}$ the index sets $I_{1}^{m}$ and
$I_{2}^{m}$ are defined as follows:
\begin{equation}
I_{1}^{m}=\{2,\ldots,n-m+1\},\qquad I_{2}^{m}=\{n-m+2,\ldots,n\},\qquad
m=0,\ldots,n+1. \label{I}%
\end{equation}
with the following degenerations for $m=0$ and for $m=n+1$:%
\[
I_{1}^{0}=I_{1}^{1}=\{2,\dots,n\}\text{, \ }I_{2}^{0}=I_{2}^{1}=\emptyset
\text{, , }I_{1}^{n+1}=I_{1}^{n}=\emptyset\text{, }I_{2}^{n+1}=I_{2}%
^{n}=\{2,\dots,n\}.
\]
With this choice of $v_{ik}$ the Hamiltonians $\mathcal{E}_{r}$ in
(\ref{qsep}) span a ($m$-dependent) Lie algebra $\mathfrak{g}=\mathrm{span}%
\{\mathcal{E}_{r}\in C^{\infty}(\mathcal{M})\colon$ $r=1,\ldots,n\}$ with the
following commutation relations \cite{mb}:
\[
\{\mathcal{E}_{1},\mathcal{E}_{r}\}=0,\quad r=2,\dots,n,
\]
and
\begin{equation}
\{\mathcal{E}_{r},\mathcal{E}_{s}\}=%
\begin{cases}
0, & \text{for $r\in I_{1}^{m}$ and $s\in I_{2}^{m}$},\\
(s-r)\mathcal{E}_{r+s-(n-m+2)}, & \text{for $r,s\in I_{1}^{m}$},\\
-(s-r)\mathcal{E}_{r+s-(n-m+2)}, & \text{for $r,s\in I_{2}^{m}$},
\end{cases}
\label{str}%
\end{equation}
Throughout the whole article we use the convention that $\mathcal{E}_{r}=0$ as
soon as $r\leq0$ or $r>n$. The algebra $\mathfrak{g}$ has an Abelian
subalgebra
\begin{equation}
\mathfrak{a}=\mathrm{span}\left\{  \mathcal{E}_{1},\dotsc,\mathcal{E}%
_{\kappa_{1}},\mathcal{E}_{n-\kappa_{2}+1},\dotsc,\mathcal{E}_{n}\right\}
\label{gma}%
\end{equation}
where%
\[
\kappa_{1}=\left[  \frac{n+3-m}{2}\right]  ,\qquad\kappa_{2}=\left[  \frac
{m}{2}\right]  .
\]
(so that $\mathfrak{a}$ depends on $m$ as well). Note that $\mathfrak{g}%
=\mathfrak{a}$ precisely when $\kappa_{1}+\kappa_{2}=n$ as $\dim
\mathfrak{a=}\kappa_{1}+\kappa_{2}$.

In what follows we will often work in the so called Vi\`{e}te canonical
coordinates
\begin{equation}
q_{i}=\rho_{i}(\lambda),\quad p_{i}=\ -\sum_{k=1}^{n}\frac{\lambda_{k}%
^{n-i}\mu_{k}}{\Delta_{k}},\quad i=1,\dotsc,n \label{26}%
\end{equation}
in which all functions $\mathcal{E}_{r}(q,p)$ are polynomial functions of
their arguments. Explicitly
\begin{align*}
G^{ij}  &  =%
\begin{cases}
q_{i+j+m-n-1}, & i,j=1,\dotsc,n-m\\
-q_{i+j+m-n-1}, & i,j=n-m+1,\dotsc,n\\
0, & \text{otherwise}%
\end{cases}
\qquad m=0,\dotsc,n\\
G^{ij}  &  =q_{i}q_{j}-q_{i+j},\quad i,j=1,\ldots,n,\quad m=n+1
\end{align*}
and
\[
\left(  K_{r}\right)  _{j}^{i}=%
\begin{cases}
q_{i-j+r-1}, & \text{$i\leq j$ and $r\leq j$}\\
-q_{i-j+r-1}, & \text{$i>j$ and $r>j$}\\
0, & \text{otherwise}%
\end{cases}
\]
where we set $q_{0}=1$, $q_{k}=0$ for $k<0$ or $k>n$. Moreover, the linear in
momenta terms in (\ref{qsep}) attain in Vi\`{e}te coordinates the form
\begin{equation}%
\begin{split}
W_{r}  &  =\sum\limits_{k=n-m-r+2}^{n-m}(n+1-m-k)q_{m+r-n-2+k}\,p_{k},\quad
r\in I_{1}^{m}\\
W_{r}  &  =-\sum\limits_{k=n-m+2}^{2n-m+2-r}(n+1-m-k)q_{m+r-n-2+k}%
\,p_{k},\quad r\in I_{2}^{m}.
\end{split}
\label{kil5}%
\end{equation}

As the Hamiltonians $\mathcal{E}_{r}$ in (\ref{qsep}) do not commute, they do
not constitute a Liouville integrable system. In particular, there is no
reason to expect that they will possess a common, multi-time solution for a
given initial data $\xi_{0}$. However, in \cite{arxiv} we found
polynomial-in-times deformations $H_{r}(t_{1},\ldots,t_{n})$ of the
Hamiltonians $\mathcal{E}_{r}$ such that the Hamiltonians $H_{r}$ satisfy the
Frobenius integrability condition
\begin{equation}
\frac{\partial H_{r}}{\partial t_{s}}-\frac{\partial H_{s}}{\partial t_{r}%
}+\{H_{r},H_{s}\}=0,\quad r,s=1,\dots,n \label{fc}%
\end{equation}
(cf. (\ref{fcg})). More specifically, the deformed Hamiltonians $H_{r}$ are
given by
\begin{align}
H_{r}  &  =\mathcal{E}_{r}-\sum_{j_{1}=1}^{n}\left(  \mathrm{ad}%
_{\mathcal{E}_{j_{1}}}\mathcal{E}_{r}\right)  t_{j_{1}}+\sum_{j_{1}=1}^{n}%
\sum_{j_{2}=j_{1}}^{n}\alpha_{rj_{1}j_{2}}\left(  \mathrm{ad}_{\mathcal{E}%
_{j_{2}}}\mathrm{ad}_{\mathcal{E}_{j_{1}}}\mathcal{E}_{r}\right)  t_{j_{1}%
}t_{j_{2}}\nonumber\\
&  \quad{}+\sum_{j_{1}=1}^{n}\sum_{j_{2}=j_{1}}^{n}\sum_{j_{3}=j_{2}}%
^{n}\alpha_{rj_{1}j_{2}j_{3}}\left(  \mathrm{ad}_{\mathcal{E}_{j_{3}}%
}\mathrm{ad}_{\mathcal{E}_{j_{2}}}\mathrm{ad}_{\mathcal{E}_{j_{1}}}%
\mathcal{E}_{r}\right)  t_{j_{1}}t_{j_{2}}t_{j_{3}}+\dotsb, \label{7}%
\end{align}
where $\mathrm{ad}_{\mathcal{E}_{i}}\mathcal{E}_{j}=\{\mathcal{E}%
_{i},\mathcal{E}_{j}\}$ and where the real constants $\alpha_{rj_{1}\cdots
j_{k}}$ can be uniquely determined from the Frobenius integrability condition
(\ref{fc}). Due to the structure of (\ref{str}) the expressions on the right
hand side of (\ref{7}) terminate. From (\ref{str}) and (\ref{gma}) it follows
that for $r\in\{1\}\cup I_{1}^{m}$
\begin{equation}
\begin{aligned} H_{r}& =\mathcal{E}_{r},\quad \text{for $r=1,\dotsc,\kappa _{1}$}, \\ H_{r}& =\sum_{j=1}^{r}\zeta _{r,j}(t_{1},\dotsc,t_{r-1})\mathcal{E}_{j},\quad \zeta _{r,r}=1,\quad \text{for $r=\kappa _{1}+1,\dotsc,n-m+1$} \end{aligned} \label{7b}%
\end{equation}
and for $r\in I_{2}^{m}$
\begin{equation}
\begin{aligned} H_{r}& =\sum_{j=0}^{n-r}\zeta _{r,r+j}(t_{r+1},\dotsc,t_{n})\mathcal{E}_{r+j},\quad \zeta _{r,r}=1,\quad \text{for $r=n-m+2,\dotsc,n-\kappa _{2}$}, \\ H_{r}& =\mathcal{E}_{r},\quad \text{for $r=n-\kappa _{2}+1,\dotsc,n$}. \end{aligned} \label{7c}%
\end{equation}
where $\zeta_{j}$ are polynomial functions that can be determined from
Frobenius conditions (\ref{fc}). For details of this construction, we refer
the reader to \cite{arxiv}, especially formula (9), Theorem 2 (formula (19)),
Corollary 1 (formula (21)) and pages 260-261 in this reference.

The functions $H_{r}$ define $n$ non-autonomous Hamiltonian systems on
$\mathcal{M}$
\begin{equation}
\xi_{t_{r}}=Y_{r}(\xi,t)=\pi dH_{r}(\xi,t),\quad r=1,\dots,n \label{nhs}%
\end{equation}
which by (\ref{fc}) are integrable in the Frobenius sense. It means that the
systems (\ref{nhs}) have a unique (local) common multi-time solution $\xi
=\xi(t_{1},\dots,t_{n},\xi_{0})$ for any initial condition $\xi_{0}$
\cite{Fecko,Lundell}. From (\ref{fc}) it follows that vector fields $Y_{r}$
satisfy the Frobenius condition (\ref{fcv}).

\begin{remark}
Due to (\ref{7}) the set of $n$ autonomous vector fields $X_{r}=\pi
d\mathcal{E}_{r}$ and the set of $n$ non-autonomous vector fields $Y_{r}=\pi
dH_{r}$ span the same distribution on $\mathcal{M}$.
\end{remark}

Another consequence of these formulas is that the\ Hamiltonians from the
Abelian subalgebra $\mathfrak{a}$ given by (\ref{gma}) remain undeformed, i.e.
$H_{r}=\mathcal{E}_{r}$ for $\mathcal{E}_{r}\in\mathfrak{a}$. In particular,
as it follows from (\ref{7b}) and (\ref{7c}), for $n=2$ we have that
$H_{r}=\mathcal{E}_{r}$ for all $r=1,2$ as then $\mathfrak{g}$ $=\mathfrak{a}%
$. For $n=3$ the formulas (\ref{7b}) and (\ref{7c}) yield that $\mathfrak{g}$
$=\mathfrak{a}$ for $m=0,2,4$ and in these cases again $H_{r}=\mathcal{E}_{r}$
for all $r$. For $n=3$ and $m=1$ we have $\mathfrak{a}=\mathrm{span}\left\{
\mathcal{E}_{1},\mathcal{E}_{2}\right\}  $ and
\begin{equation}
H_{r}=\mathcal{E}_{r}\text{ for $r=1,2$}, \quad H_{3}=\mathcal{E}_{3}%
+t_{2}\mathcal{E}_{1} \label{31}%
\end{equation}
while for $n=3$ and $m=3$ we have $\mathfrak{a}=\mathrm{span}\left\{
\mathcal{E}_{1},\mathcal{E}_{3}\right\}  $ and%
\begin{equation}
H_{r}=\mathcal{E}_{r}\text{ for $r=1,3$},\quad H_{2}=\mathcal{E}_{2}%
+t_{3}\mathcal{E}_{3}. \label{33}%
\end{equation}

\begin{example}
Let us now present a higher dimensional example when $n=11$, $m=6$. Then
$\kappa_{1}=4$, $\kappa_{2}=3$ and $\mathfrak{a=}\mathrm{span}\left\{
\mathcal{E}_{1},\ldots,\mathcal{E}_{4},\mathcal{E}_{9},\ldots,\mathcal{E}%
_{11}\right\}  $. From (\ref{7b}) and (\ref{7c}) it follows that the deformed
Hamiltonians are given by
\begin{align*}
H_{r}  &  =\mathcal{E}_{r},\quad r=1,\dots,4,9,\ldots,11,\\
H_{5}  &  =\mathcal{E}_{5}+t_{4}\mathcal{E}_{2}+2t_{3}\mathcal{E}_{1},\\
H_{6}  &  =\mathcal{E}_{6}+4t_{2}\mathcal{E}_{1}+(3t_{3}-\tfrac{1}{2}t_{5}%
^{2})\mathcal{E}_{2}+2t_{4}\mathcal{E}_{3}+t_{5}\mathcal{E}_{4},\\
H_{7}  &  =\mathcal{E}_{7}+t_{8}\mathcal{E}_{8}+2t_{9}\mathcal{E}_{9}%
+(3t_{10}+t_{8}t_{9})\mathcal{E}_{10}+(4t_{11}+2t_{8}t_{10})\mathcal{E}%
_{11},\\
H_{8}  &  =\mathcal{E}_{8}+t_{9}\mathcal{E}_{10}+2t_{10}\mathcal{E}_{11}.
\end{align*}
and one can verify that the Hamiltonians $H_{1},\dotsc,H_{11}$ do satisfy the
Frobenius condition (\ref{fc}).
\end{example}

Finally, let us remark that the non-autonomous Hamiltonian equations
(\ref{nhs}) are conservative, as by (\ref{7b})--(\ref{7c}) the $r$-th
Hamiltonian $H_{r}$ does not depend in explicit way on its own evolution
parameter $t_{r} $.

\section{Frobenius integrable deformations of quasi-St\"{a}ckel systems with
potentials\label{sec4}}

In this section we are going to construct, in a systematic way,
multi-parameter families of Frobenius integrable non-autonomous Hamiltonian
systems with potentials. We will achieve it through appropriate
(multi)time-dependent deformations of quasi-St\"{a}ckel systems with
potentials. Let us thus consider the following quasi-separation relations
\begin{equation}
\sum_{\alpha\in A}c_{\alpha}(t_{1},\dotsc,t_{n})\lambda_{i}^{\alpha}%
+\sum_{r=1}^{n}\lambda_{i}^{n-r}h_{r}^{A}=\frac{1}{2}\lambda_{i}^{m}\mu
_{i}^{2}+\sum_{k=1}^{n}v_{ik}(\lambda)\mu_{k},\quad i=1,\ldots,n, \label{sepg}%
\end{equation}
where as before $m\in\{0,\dots,n+1\}$, $A\subset\mathbb{Z}$ is a finite subset
of integers and where $v_{ik}$ are again given by (\ref{ni}). The system
(\ref{sepg}) naturally generalizes (\ref{sep}). Solving this system with
respect to $h_{r}^{A}$ we obtain%
\begin{equation}
h_{r}^{A}=\mathcal{E}_{r}+V_{r}^{A}=E_{r}+W_{r}+V_{r}^{A},\quad r=1,\ldots,n
\label{hg1}%
\end{equation}
where $E_{r}$, $W_{r}$ are given respectively by (\ref{4}) and (\ref{kil5}).
The functions%
\[
V_{r}^{A}=\sum_{\alpha\in A}c_{\alpha}(t_{1},\dotsc,t_{n})V_{r}^{(\alpha)}%
\]
on the base manifold $Q$ are time-dependent linear combinations of the so
called basic (elementary) separable potentials $V_{r}^{(\alpha)}$. By
linearity of (\ref{sepg}), the potentials $V_{r}^{(\alpha)}$ satisfy the
relations%
\[
\lambda_{i}^{\alpha}+\sum_{r=1}^{n}V_{r}^{(\alpha)}\lambda_{i}^{n-r}=0,\quad
i=1,\dotsc,n,\quad\alpha\in\mathbb{Z},
\]
(we stress that they do not depend on $m$) so, by Lemma \ref{VDM}, they are
given by
\begin{equation}
V_{r}^{(\alpha)}=\sum_{i=1}^{n}\frac{\partial\rho_{r}}{\partial\lambda_{i}%
}\frac{\lambda_{i}^{\alpha}}{\Delta_{i}},\quad r=1,\ldots,n \label{Vnew}%
\end{equation}
and can be explicitly constructed by the recursion formula \cite{blasz2011}
\begin{equation}
V^{(\alpha)}=R^{\alpha}V^{(0)},\qquad V^{(\alpha)}=(V_{1}^{(\alpha)}%
,\dotsc,V_{n}^{(\alpha)})^{T}, \label{6}%
\end{equation}
where in Vi\`{e}te coordinates
\begin{equation}
R=%
\begin{pmatrix}
-q_{1} & \ 1 & \ 0 & \ 0\\
\vdots & \ 0 & \ \ddots & \ 0\\
\vdots & \ 0 & \ 0 & \ 1\\
-q_{n} & \ 0 & \ 0 & \ 0
\end{pmatrix}
\label{6a}%
\end{equation}
with $V^{(0)}=(0,\dotsc,0,-1)^{T}$. The formulas (\ref{6})--(\ref{6a}) are
non-tensorial in the sense that they are true in any coordinate system on the
base manifold $Q$. The first $n$ basic separable potentials are trivial%
\begin{equation}
V_{k}^{(\alpha)}=-\delta_{k,n-\alpha},\quad\alpha=0,\dotsc,n-1. \label{7a}%
\end{equation}
The first two nontrivial positive potentials are
\begin{align*}
V^{(n)}  &  =(q_{1},\dotsc,q_{n})^{T}\\
V^{(n+1)}  &  =(-q_{1}^{2}q_{2},-q_{1}q_{2}+q_{3},-q_{1}q_{3}+q_{4}%
,\ldots,-q_{1}q_{n-1}+q_{n},-q_{1}q_{n})^{T}%
\end{align*}
and higher potentials are more complicated polynomials in $q_{i}$. The first
two negative potential are
\begin{align*}
V^{(-1)}  &  =\left(  \frac{1}{q_{n}},\frac{q_{1}}{q_{n}},\dotsc,\frac
{q_{n-1}}{q_{n}}\right)  ^{T}\\
V^{(-2)}  &  =\left(  -\frac{q_{n-1}}{q_{n}^{2}},\frac{1}{q_{n}}-\frac
{q_{1}q_{n-1}}{q_{n}^{2}},\frac{q_{1}}{q_{n}}-\frac{q_{2}q_{n-1}}{q_{n}^{2}%
},\frac{q_{2}}{q_{n}}-\frac{q_{3}q_{n-1}}{q_{n}^{2}},\dotsc,\frac{q_{n-2}%
}{q_{n}}-\frac{q_{n-1}^{2}}{q_{n}^{2}}\right)  ^{T}%
\end{align*}
and the higher negative potentials are more complicated rational functions of
all $q_{i}$.

In this section, we will solve the following problem.

\begin{problem}
\label{problem}For an arbitrary $m\in\{0,\dots,n+1\}$, determine the set $A$
as well as the explicit form of the coefficients $c_{\alpha}(t_{1}%
,\dotsc,t_{n})$ such that the system (\ref{sepg}), can be deformed by formulas
(\ref{7b})--(\ref{7c}) to a Frobenius integrable non-autonomous system
satisfying (\ref{fc}).
\end{problem}

In order to solve this problem we will first consider Hamiltonians (\ref{hg1})
containing a \emph{single} basic separable potential $V_{r}^{(\alpha)}$ with a
fixed $\alpha\in\mathbb{Z}$
\begin{equation}
h_{r}^{(\alpha)}=\mathcal{E}_{r}+V_{r}^{(\alpha)}=E_{r}+W_{r}+V_{r}^{(\alpha
)},\quad r=1,\ldots,n. \label{hg2}%
\end{equation}
The following commutation relations are valid between the Hamiltonians
$h_{r}^{(\alpha)}$ (we focus only on cases when $\alpha\geq n$ or $\alpha<0 $
as otherwise the potentials $V_{r}^{(\alpha)}$ are trivial and the system
becomes simply the geodesic quasi-St\"{a}ckel system, analyzed in the previous section):

\begin{theorem}
\label{1t} Consider the Hamiltonians (\ref{hg2}). The following commutation
relations hold:

\begin{description}
\item[(i)] when $r,s\in\left\{  1\right\}  \cup I_{1}^{m}$
\begin{multline}
\left\{  h_{r}^{(n+k)},h_{s}^{(n+k)}\right\}  = (s-r)h_{r+s+m-n-2}^{(n+k)}\\
+(2r+k+m-n-2)V_{s}^{(r+k+m-2)}-(2s+k+m-n-2)V_{r}^{(s+k+m-2)} \label{t1}%
\end{multline}
for $k=0,\dotsc,n-m+2$ and
\begin{equation}
\left\{  h_{r}^{(-k)},h_{s}^{(-k)}\right\}  =(s-r)h_{r+s+m-n-2}^{(-k)}
\label{t2}%
\end{equation}
for $k=1,\dotsc,m$,

\item[(ii)] when $r,s\in I_{2}^{m}$%
\begin{equation}
\left\{  h_{r}^{(n+k)},h_{s}^{(n+k)}\right\}  =(r-s)h_{r+s+m-n-2}^{(n+k)}
\label{t3}%
\end{equation}
for $k=0,\dotsc,n-m+2$ and
\begin{multline}
\left\{  h_{r}^{(-k)},h_{s}^{(-k)}\right\}  = (r-s)h_{r+s+m-n-2}^{(-k)}\\
+(2s-k+m-2n-2)V_{r}^{(s-k+m-n-2)}-(2r-k+m-2n-2)V_{s}^{(r-k+m-n-2)} \label{t4}%
\end{multline}
for $k=1,\dotsc,m$,

\item[(iii)] when $r\in\left\{  1\right\}  \cup I_{1}^{m}$, $s\in I_{2}^{m}$%
\begin{equation}
\left\{  h_{r}^{(n+k)},h_{s}^{(n+k)}\right\}  =(2r+k+m-n-2)V_{s}^{(k+r+m-2)}
\label{t5}%
\end{equation}
for $k=0,\dotsc,n-m+2$ and
\begin{equation}
\left\{  h_{r}^{(-k)},h_{s}^{(-k)}\right\}  =(2s-k+m-2n-2)V_{r}^{(-k+s+m-n-2)}
\label{t6}%
\end{equation}
for $k=1,\dotsc,m$.
\end{description}
\end{theorem}

For other combinations of indices $r,s,m$ and $k$, the Poisson bracket of two
Hamiltonians $h_{r}^{(\alpha)}$ is not a linear combination of a Hamiltonian
$h_{j}^{(\alpha)}$ and basic separable potentials.The proof of this theorem
can be found in Appendix~A.

As we see, for a given $\alpha=n+k\in$ $\left\{  n,\ldots,2n-m+2\right\}  $,
the additional potentials $V_{r}^{(\delta)}$ on the right hand sides of
(\ref{t1}) and (\ref{t5}) are such that $\delta\in\left\{  n,\ldots
,2n-m+2\right\}  $ as well. The same is true for $\alpha=-k\in\left\{
-m,\ldots,-1\right\}  $ in (\ref{t4}) and (\ref{t6}). This leads to the
conclusion, that for a given $m\in\{0,\dots,n+1\}$ the set $A$ must be of the
form%
\[
A=A_{\tau_{1}\tau_{2}}=\left\{  \tau_{1},\ldots,\tau_{2}\right\}  , \quad
\tau_{1}\geq-m,\ \tau_{2}\leq2n-m+2,\ \tau_{1}\leq\tau_{2}.
\]
Having chosen $\tau_{1}\geq-m$, $\tau_{2}\leq2n-m+2$ we obtain $h_{r}^{A} $ in
(\ref{hg1}) of the form
\begin{equation}
h_{r}^{A}=\mathcal{E}_{r}+V_{r}^{A}=E_{r}+W_{r}+\sum_{\alpha=\tau_{1}}%
^{\tau_{2}}c_{\alpha}(t_{1},\dotsc,t_{n})V_{r}^{(\alpha)}, \quad r=1,\ldots,n.
\label{jeden}%
\end{equation}
We can now establish the sought functions $c_{\alpha}$, $\alpha=\tau
_{1},\ldots,\tau_{2}$, using the following procedure.

\begin{enumerate}
\item We deform the Hamiltonians $h_{r}^{A}$ given by (\ref{jeden}) through
the formulas (\ref{7b}) and (\ref{7c}) to Hamiltonians $H_{r}^{A}$.

\item We impose the Frobenius condition (\ref{fc}) on the Hamiltonians
$H_{r}^{A}$ which leads to a complicated system of first order PDE's for the
unknown functions $\zeta_{r,j}(t_{1},\dotsc,t_{r-1}),$ $\zeta_{r,r+j}%
(t_{r+1},\dotsc,t_{n})$ and all $c_{\alpha}(t_{1},\dotsc,t_{n})$. This system
contains a subsystem not involving $c_{\alpha}(t_{1},\ldots,t_{n})$, that is
identical to the system that originates during deforming the geodesic
quasi-St\"{a}ckel system and the remaining part, that also involves
$c_{\alpha}(t_{1},\ldots,t_{n})$.

\item The subsystem not involving $c_{\alpha}(t_{1},\ldots,t_{n})\,\ $yields a
unique solution on the functions $\zeta_{r,j}(t_{1},\dotsc,t_{r-1}),$
$\zeta_{r,r+j}(t_{r+1},\dotsc,t_{n})$ (provided that we choose all the
integration constants equal to zero). These solutions are exactly the same as
calculated for geodesic quasi-St\"{a}ckel systems in Section 3.

\item Finally, we find the explicit form of functions $c_{\alpha}(t_{1}%
,\ldots,t_{n})$, recursively solving the remaining part of the system.
\end{enumerate}

Note that the potential in (\ref{jeden}) contains in general also trivial
separable potentials (for $\alpha$ between $0$ and $n-1$) that do not
influence the dynamics of the systems. The system of PDE's obtained in the
last step is underdetermined in the functions $c_{\alpha}$ for $\alpha
\in\left\{  0,\ldots,n-1\right\}  $. Note also that the maximal possible set
$A_{\tau_{1}\tau_{2}}$ is $A=\left\{  -m,\ldots,2n-m+2\right\}  $. Le us
illustrate these statements through the following example.

\begin{example}
\label{1e}Consider the case $n=3$, $m=1$ and $\tau_{1}=0,\tau_{2}=5$ so that
$A=A_{\tau_{1}\tau_{2}}=\left\{  0,\ldots,5\right\}  $. Then, the
quasi-St\"{a}ckel Hamiltonians $h_{r}^{A}$ in (\ref{jeden}) are%
\[
h_{r}^{A}=\mathcal{E}_{r}+\sum_{\alpha=0}^{5}c_{\alpha}(t_{1},\dotsc
,t_{n})V_{r}^{(\alpha)}\text{, }r=1,2,3\text{,}%
\]
where in Vi\`{e}te coordinates
\begin{equation}
\begin{aligned} \mathcal{E}_{1}&= p_{{1}}p_{{2}}+\frac{1}{2}\,q_{{1}}{p}_{2}^{2}-\frac{1}{2}\,{p_{{3}}}^{2}q_{{3}}, \\ \mathcal{E}_{2}&= p_{{2}}q_{{1}}p_{{1}}+\frac{1}{2}{p}_{1}^{2}-q_{{3}}p_{{2}}p_{{3}}+\frac{1}{2}\left( \,{q}_{1}^{2}-\,q_{{2}}\right) {p_{{2}}}^{2}-\frac{1}{2}\,q_{{1}}q_{{3}}{p}_{3}^{2}+p_{2}, \\ \mathcal{E}_{3}&= -q_{{3}}p_{{1}}p_{{3}}-\frac{1}{2}\,q_{{3}}{p}_{2}^{2}-q_{{1}}q_{{3}}p_{{2}}p_{{3}}-\frac{1}{2}\,q_{{2}}q_{{3}}{p}_{3}^{2}+q_{1}p_{2}+2p_{1} \end{aligned} \label{E1a}%
\end{equation}
while $V_{r}^{(\alpha)}$ are given by formulas (\ref{Vnew}) or by (\ref{6}).
For example
\[
V_{1}^{(5)}={q_{{1}}^{3}}-2\,q_{{1}}q_{{2}}+{q_{{3}}},\quad V_{2}%
^{(5)}={q_{{1}}^{2}}q_{{2}}-\,q_{{1}}{q_{{3}}}-\,q_{{2}}^{2},\quad V_{3}%
^{(5)}={q_{{1}}^{2}}q_{{3}}-q_{{2}}q_{{3}}.
\]
As $\kappa_{1}+\kappa_{2}<n$ we have to perform\ the first three steps of our
procedure, resulting in the deformation of the Hamiltonians $h_{r}^{A}$ to the
Hamiltonians $H_{r}^{A}$, exactly as given by (\ref{31}):%
\begin{equation}
H_{1}^{A}=h_{1}^{A},\quad H_{2}^{A}=h_{2}^{A},\quad H_{3}^{A}=h_{3}^{A}%
+t_{2}h_{1}^{A}. \label{E3}%
\end{equation}
In the last step we find the functions $c_{\alpha}$ from the Frobenius
condition (\ref{fc}). Inserting (\ref{E3}) into the left hand side of
(\ref{fc}) we obtain a compatible underdetermined (in the functions $c_{0}%
$,$c_{1},c_{2}$) set of first order PDE's on $c_{\alpha}$ that can be solved
recursively, starting from $c_{5}$. The set of PDE's for $c_{5}$ is
$\frac{\partial c_{5}}{\partial t_{i}}=0$ for $i=1,2,3$ so that $c_{5}%
(t)=a_{5}\in\mathbb{R}$. Further we obtain
\[
\frac{\partial c_{4}}{\partial t_{1}}=0,\quad\frac{\partial c_{4}}{\partial
t_{2}}=0,\quad\frac{\partial c_{4}}{\partial t_{3}}=4a_{5}%
\]
and integrating it we obtain $c_{4}=4a_{5}t_{3}+a_{4}$. Plugging this solution
(with $a_{4}$ chosen to be $0$ for simplicity) again to our set of PDE's we
obtain the set of PDE's for $c_{3}$%
\[
\frac{\partial c_{3}}{\partial t_{1}}=0,\quad\frac{\partial c_{3}}{\partial
t_{2}}=2a_{5},\quad\frac{\partial c_{3}}{\partial t_{3}}=12a_{5}t_{3}%
\]
with the solution $c_{3}=2a_{5}(3t_{3}^{2}+t_{2})+a_{3}$ where we again choose
$a_{3}=0$. With these solutions found, the remaining PDE's for $c_{0},c_{1}$
and $c_{2}$ attain the under-determined form
\begin{equation}
\begin{aligned} 4a_{5}t_{3}-\frac{\partial c_{2}}{\partial t_{2}}+\frac{\partial c_{1}}{\partial t_{1}}& =0, \\ 12a_{5}t_{3}^{2}+4a_{5}t_{2}-\frac{\partial c_{2}}{\partial t_{3}}+t_{2}\frac{\partial c_{2}}{\partial t_{1}}+\frac{\partial c_{0}}{\partial t_{1}}& =0, \\ -4a_{5}t_{2}t_{3}+c_{2}+\text{ }\frac{\partial c_{0}}{\partial t_{2}}-\frac{\partial c_{1}}{\partial t_{3}}+\frac{\partial c_{2}}{\partial t_{2}}t_{2}& =0. \end{aligned} \label{sol}%
\end{equation}
If for example we choose $c_{0}=0$, then the remaining PDE's yield
$c_{1}=a_{5}(t_{3}^{3}-2t_{2}t_{3}-4t_{1})t_{3},$ $c_{2}=4a_{5}(t_{3}%
^{3}-t_{1})$.
\end{example}

In general our procedure leads to a $(n+3)$-parameter family of Frobenius
integrable non-autonomous systems with potentials. Although the obtained
systems are parametrized by $2n+3$ integration constants $a_{-m},\ldots,
a_{2n-m+2}$, the $n$ constants $a_{0},\ldots,a_{n-1}$ are integration
constants that originate in the trivial potentials $V_{r}^{(0)},\ldots
,V_{r}^{(n-1)}$ and as such enter the Hamiltonians only in a trivial way,
through some undetermined functions of times only, not affecting the dynamics
of the systems. We can thus say that our systems are parametrized by $n+3$
dynamical parameters $\left(  a_{-m},\ldots,a_{-1},a_{n},\dotsc,a_{2n-m+2}%
\right)  $ and by $n$ non-dynamical parameters $\left(  a_{0},\dotsc
,a_{n-1}\right)  $.

\section{Frobenius integrable deformations of two- and three-dimensional
quasi-St\"{a}ckel systems with potentials\label{sec5}}

In this section we present a complete list of Frobenius integrable (satisfying
(\ref{fc})) deformations (\ref{7b}), (\ref{7c}) of two- and three-dimensional
quasi-St\"{a}ckel systems with potentials that originate in our deformation
procedure. We present here only the results and we always make use of the
maximal set $A=\left\{  -m,\ldots,2n-m+2\right\}  $. Each obtained Hamiltonian
$H_{r}^{A}$ is determined up to a function (given by a compatible but
under-determined set of PDE's) of $t_{1},\ldots,t_{n}$ and of the non-dynamic
integration constants $a_{0},\ldots,a_{n-1}$. The basic separable potentials
in the formulas below are given by (\ref{Vnew}) or equivalently by (\ref{6}),
(\ref{6a}). In each case $(n,m)$ we obtain a $(2n+3)$-parameter family of
systems, parametrized by $n+3$ dynamical constants $a_{-m},\ldots,a_{-1}%
,a_{n},\dotsc,a_{2n-m+2}$ and by $n$ non-dynamic integration constants
$a_{0},\ldots,a_{n-1}$.

\subsection{Two-dimensional systems}

Let us first consider the case $n=2$ so that $A=\left\{  -m,\ldots
,6-m\right\}  $. As it was explained in Section \ref{bla} in this case
$H_{r}^{A}=h_{r}^{A}$ in (\ref{7b}), (\ref{7c}) and for each $m=0,\dotsc,3$ we
obtain a $2n+3=7$-parameter family of Frobenius integrable non-autonomous
Hamiltonian systems satisfying (\ref{fc}), with two non-dynamical parameters
$(a_{0},a_{1})$ an five dynamical parameters $a=(a_{-m},\dotsc,a_{-1}%
,a_{2},\dotsc,a_{6-m})$.

For $m=0$ we have $A=\left\{  0,\ldots,6\,\right\}  $ and explicitly, in
Vi\`{e}te coordinates we get
\begin{align}
h_{r}^{A}  &  =\mathcal{E}_{r}+a_{6}V_{r}^{(6)}+a_{5}V_{r}^{(5)}+(a_{4}%
+4a_{6}t_{2})V_{r}^{(4)}+(a_{3}+3a_{5}t_{2}+2a_{6}t_{1})V_{r}^{(3)}\nonumber\\
&  \quad{}+(a_{2}+2a_{4}t_{2}+a_{5}t_{1}+4a_{6}t_{2}^{2})V_{r}^{(2)}%
-c_{n-r}(t_{1},t_{2,}a_{0,}a_{1},a) \label{e1}%
\end{align}
(the form of the non-dynamic part follows from (\ref{sepg})) where
\begin{equation}
\mathcal{E}_{1}=p_{{1}}p_{{2}}+\frac{1}{2}q_{{1}}\,{p_{{2}}^{2}}%
,\qquad\mathcal{E}_{2}=q_{{1}}p_{{1}}p_{{2}}+\frac{1}{2}\,{p_{{1}}^{2}}%
+\frac{1}{2}\left(  {q_{{1}}^{2}}-q_{{2}}\right)  {p_{{2}}^{2}+p}_{2}.
\label{geo1}%
\end{equation}

For $m=1$ we have $A=\left\{  -1,\ldots,5\right\}  $ and we get
\begin{align}
h_{r}^{A}  &  =\mathcal{E}_{r}+a_{5}V_{r}^{(5)}+(a_{4}+4a_{5}t_{2})V_{r}%
^{(4)}+[a_{3}+3a_{4}t_{2}+2a_{5}(t_{1}+3t_{2}^{2})]V_{r}^{(3)}\nonumber\\
&  \quad{} +[a_{2}+2a_{3}t_{2}+a_{4}(t_{1}+3t_{2}^{2})+4a_{5}(t_{1}t_{2}%
+t_{2}^{3})]V_{r}^{(2)}+a_{-1}V_{r}^{(-1)}-c_{n-r}(t_{1},t_{2,}a_{0,}a_{1},a)
\label{e2}%
\end{align}
where
\begin{equation}
\mathcal{E}_{1}=\frac{1}{2}\,{p}_{1}^{2}-\frac{1}{2}\,q_{{2}}{p}_{2}%
^{2},\qquad\mathcal{E}_{2}=-q_{{2}}p_{{2}}p_{{1}}-\frac{1}{2}\,q_{{1}}q_{{2}%
}{p}_{2}^{2}+p_{1}. \label{geo2}%
\end{equation}
For $m=2$ we have $A=\left\{  -2,\ldots,4\right\}  $ and
\begin{align}
h_{r}^{A}  &  =\mathcal{E}_{r}+a_{4}V_{r}^{(4)}+(a_{3}+2a_{4}t_{1})V_{r}%
^{(3)}+(a_{2}+a_{3}t_{1}+a_{4}t_{1}^{2})V_{r}^{(2)}\nonumber\\
&  \quad{} +a_{-1}{\mathrm{e}^{t_{{2}}}}V_{r}^{(-1)}+a_{-2}{\mathrm{e}%
^{2t_{{2}}}V}_{r}^{(-2)}-c_{n-r}(t_{1},t_{2,}a_{0,}a_{1},a) \label{e3}%
\end{align}
where
\begin{equation}
\mathcal{E}_{1}=-\frac{1}{2}q_{{1}}\,{p}_{1}^{2}-q_{{2}}p_{{2}}p_{{1}}%
,\qquad\mathcal{E}_{2}=-\frac{1}{2}\,q_{{2}}{p}_{1}^{2}+\frac{1}{2}{q}_{2}%
^{2}\,{p}_{2}^{2}+q_{{2}}p_{{2}}. \label{geo3}%
\end{equation}
For $m=3$ we have $A=\left\{  -3,\ldots,3\right\}  $ and
\begin{align}
h_{r}^{A}  &  =\mathcal{E}_{r}+a_{3}{\mathrm{e}^{2t_{{1}}}}V_{r}^{(3)}%
+a_{2}{\mathrm{e}^{t_{{1}}}}V_{r}^{(2)}+(a_{-1}+a_{-2}t_{2}+a_{-3}t_{2}%
^{2})V_{r}^{(-1)}\nonumber\\
&  \quad{} +(a_{-2}+2a_{-3}t_{2})V_{r}^{(-2)}+a_{-3}{V}_{r}^{(-3)}%
-c_{n-r}(t_{1},t_{2,}a_{0,}a_{1},a) \label{e4}%
\end{align}
where
\begin{equation}
\mathcal{E}_{1}=\frac{1}{2}\left(  {q}_{1}^{2}-q_{{2}}\right)  {p}_{1}%
^{2}+\frac{1}{2}{q}_{2}^{2}\,{p}_{2}^{2}+q_{{1}}q_{{2}}p_{{1}}p_{{2}}%
,\qquad\mathcal{E}_{2}={q}_{2}^{2}p_{{1}}p_{{2}}+\frac{1}{2}\,q_{{1}}q_{{2}%
}{p}_{1}^{2}+q_{{2}}p_{{1}}. \label{geo4}%
\end{equation}

As the metric $G$ is flat for $m=0,1,2$, for these cases we can express all
the formulas in flat coordinates $(x,y)$ (see for example \cite{blasz2007}).

\subsection{Three-dimensional systems}

Consider now the case $n=3$ so that $A=\left\{  -m,\ldots,8-m\right\}  $. For
each $m=0,\dotsc,4$ we obtain a $9$-parameter family with $3$ non-dynamical
parameters $(a_{0},a_{1},a_{2})$ and $6$ dynamical parameters $a=(a_{-m}%
,\dotsc,a_{-1},a_{2},\dotsc,a_{8-m})$, of Frobenius integrable non-autonomous
Hamiltonian systems satisfying (\ref{fc}).

For $m=0$ we have $\mathfrak{a}$ $=\mathfrak{g}$ so $H_{r}^{A}=h_{r}^{A}$ for
all $r$ and our procedure yields
\begin{align}
h_{r}^{A}  &  =\mathcal{E}_{r}+a_{{8}}V_{r}^{(8)}+a_{{7}}V_{r}^{(7)}+\left(
6\,a_{{8}}t_{{3}}+a_{{6}}\right)  V_{r}^{(6)}+\left(  5\,a_{{7}}t_{{3}%
}+4\,a_{{8}}t_{{2}}+a_{{5}}\right)  V_{r}^{(5)}\nonumber\\
&  \quad{}+[a_{{4}}+2\,a_{{8}}\left(  6\,{t}_{3}^{2}+t_{{1}}\right)
+3\,a_{{7}}t_{{2}}+4\,a_{{6}}t_{{3}}]V_{r}^{(4)}\nonumber\\
&  \quad{}+[a_{{3}}+12\,a_{{8}}t_{{2}}t_{{3}}+a_{{7}}(t_{{1}}+\tfrac{15}%
{2}\,{t}_{3}^{2})+2\,a_{{6}}t_{{2}}+3\,a_{{5}}t_{{3}}]V_{r}^{(3)}%
-c_{n-r}(t_{1},t_{2},t_{3},a_{0},a_{1},a_{2},a) \label{3.0}%
\end{align}
where in Vi\`{e}te coordinates
\begin{equation}
\begin{aligned} \mathcal{E}_{1}&=p_{{1}}p_{{3}}+\frac{1}{2}\,{p}_{2}^{2}+p_{{3}}q_{{1}}p_{{2}}+\frac{1}{2}\,q_{{2}}{p}_{3}^{2}, \\ \mathcal{E}_{2}&=q_{{1}}p_{{1}}p_{{3}}+q_{{1}}{p}_{2}^{2}+p_{{1}}p_{{2}}+\frac{1}{2}\left( q_{{1}}q_{{2}}-\,q_{{3}}\right) {p}_{3}^{2}+{q}_{1}^{2}p_{{2}}p_{{3}}+p_{3}, \\ \mathcal{E}_{3}&=q_{{2}}p_{{1}}p_{{3}}+\frac{1}{2}{q}_{1}^{2}{p}_{2}^{2}+q_{{1}}p_{{1}}p_{{2}}+\frac{1}{2}\,{p}_{1}^{2}+\frac{1}{2}\left( \,{q}_{2}^{2}-\,q_{{3}}q_{{1}}\right) {p}_{3}^{2} \\ & \quad {} +\left( q_{{1}}q_{{2}}-q_{{3}}\right) p_{{2}}p_{{3}}+2p_{2}+q_{1}p_{3}. \end{aligned} \label{geo5}%
\end{equation}

For $m=1$ we have, by (\ref{31}), $H_{1}^{A}=h_{1}^{A},~H_{2}^{A}=h_{2}^{A}$
and $H_{3}^{A}=h_{3}^{A}+t_{2}h_{1}^{A}$ where
\begin{align}
h_{r}^{A}  &  =\mathcal{E}_{r}+a_{{7}}V_{r}^{(7)}+(a_{{6}}+6a_{7}t_{3}%
)V_{r}^{(6)}+[a_{{5}}+5a_{6}t_{3}+\,a_{{7}}(4t_{{2}}+15t_{3}^{2})]V_{r}%
^{(5)}\nonumber\\
&  \quad{} +[a_{4}+4\,a_{{5}}t_{{3}}+\,a_{{6}}(3t_{{2}}+10t_{3}^{2}%
)+2a_{7}(t_{1}+9t_{2}t_{3}+10t_{3}^{3})]V_{r}^{(4)}\nonumber\\
&  \quad{} +[a_{{3}}+3\,a_{{4}}t_{{3}}+2\,a_{{5}}(t_{{2}}+3t_{3}^{2})+a_{{6}%
}\left(  t_{{1}}+10t_{2}t_{3}+10t_{3}^{3}\right) \nonumber\\
&  \quad{} +a_{7}(4t_{2}^{2}+6t_{1}t_{3}+30t_{2}t_{3}^{2}+15t_{3}^{4}%
)]V_{r}^{(3)}+a_{-1}V_{r}^{(-1)}-c_{n-r}(t_{1},t_{2},t_{3},a_{0},a_{1}%
,a_{2},a) \label{3.1}%
\end{align}
and
\begin{equation}
\begin{aligned} \mathcal{E}_{1}&=p_{{1}}p_{{2}}+\frac{1}{2}\,q_{{1}}{p}_{2}^{2}-\frac{1}{2}\,{p_{{3}}}^{2}q_{{3}}, \\ \mathcal{E}_{2}&=p_{{2}}q_{{1}}p_{{1}}+\frac{1}{2}{p}_{1}^{2}-q_{{3}}p_{{2}}p_{{3}}+\frac{1}{2}\left( \,{q}_{1}^{2}-\,q_{{2}}\right) {p_{{2}}}^{2}-\frac{1}{2}\,q_{{1}}q_{{3}}{p}_{3}^{2}+p_{2}, \\ \mathcal{E}_{3}&=-q_{{3}}p_{{1}}p_{{3}}-\frac{1}{2}\,q_{{3}}{p}_{2}^{2}-q_{{1}}q_{{3}}p_{{2}}p_{{3}}-\frac{1}{2}\,q_{{2}}q_{{3}}{p}_{3}^{2}+q_{1}p_{2}+2p_{1}. \end{aligned} \label{geo6}%
\end{equation}
Note that the system from Example \ref{1e} is a particular case of the system
(\ref{3.1}) obtained by putting $a_{-1}=a_{3}=a_{4}=a_{6}=a_{7}=0$.

For $m=2$ we have $\mathfrak{a}$ $=\mathfrak{g}$ so $H_{r}^{A}=h_{r}^{A}$ for
all $r$ with
\begin{align*}
h_{r}^{A}  &  =\mathcal{E}_{r}+a_{{6}}V_{r}^{(6)}+(a_{{5}}+4a_{6}t_{2}%
)V_{r}^{(5)}+[a_{{4}}+3a_{5}t_{2}+\,2a_{{6}}(t_{{1}}+3t_{2}^{2})]V_{r}^{(4)}\\
&  \quad{} +[a_{3}+2a_{{4}}t_{{2}}+\,a_{{5}}(t_{{1}}+3t_{2}^{2})+4a_{6}%
(t_{1}t_{2}+t_{2}^{3})]V_{r}^{(3)}+a_{-1}{\mathrm{e}^{t_{{3}}}V}_{r}%
^{(-1)}+a_{-2}{\mathrm{e}^{2t_{{3}}}V}_{r}^{(-2)}\\
&  \quad{} -c_{n-r}(t_{1},t_{2},t_{3},a_{0},a_{1},a_{2},a),
\end{align*}
where
\begin{equation}
\begin{aligned} \mathcal{E}_{1}&=\frac{1}{2}\,{p}_{1}^{2}-\frac{1}{2}q_{{2}}\,{p}_{2}^{2}-q_{{3}}p_{{2}}p_{{3}}, \\ \mathcal{E}_{2}&=-q_{{2}}p_{{1}}p_{{2}}-q_{{3}}p_{{1}}p_{{3}}-q_{{1}}q_{{3}}p_{{2}}p_{{3}}-\frac{1}{2}\left( q_{{1}}q_{{2}}+\,q_{{3}}\right) {p}_{2}^{2}+p_{1}, \\ \mathcal{E}_{3}&=-q_{{3}}p_{{1}}p_{{2}}-\frac{1}{2}\,q_{{1}}q_{{3}}{p}_{2}^{2}+\frac{1}{2}\,{q}_{3}^{2}{p}_{3}^{2}+q_{3}p_{3}. \end{aligned} \label{geo7}%
\end{equation}

For $m=3$ we have, due to (\ref{33}), $H_{1}^{A}=h_{1}^{A},~H_{3}^{A}%
=h_{3}^{A}$ and $H_{2}^{A}=h_{2}^{A}+t_{3}h_{3}^{A}$ with
\begin{align*}
h_{r}^{A}  &  =\mathcal{E}_{r}+a_{{5}}V_{r}^{(5)}+(a_{{4}}+2a_{5}t_{1}%
)V_{r}^{(4)}+(a_{{3}}+a_{4}t_{1}+\,a_{{5}}t_{{1}}^{2})V_{r}^{(3)}\\
&  \quad{} +[a_{-1}{\mathrm{e}^{2t_{{2}}}+a}_{-2}({\mathrm{e}^{2t_{{2}}}%
+t}_{3}{\mathrm{e}^{3t_{{2}}})+a}_{-3}t_{3}^{2}{\mathrm{e}^{4t_{{2}}}]V}%
_{r}^{(-1)}+(a_{-2}{\mathrm{e}^{3t_{{2}}}+2a}_{-3}{\mathrm{e}^{4t_{{2}}}%
)V}_{r}^{(-2)}\\
&  \quad{} +a_{-3}{\mathrm{e}^{4t_{{2}}}V}_{r}^{(-3)}-c_{n-r}(t_{1}%
,t_{2},t_{3},a_{0},a_{1},a_{2},a),
\end{align*}
where
\begin{equation}
\begin{aligned} \mathcal{E}_{1}&=-\frac{1}{2}\,q_{{1}}{p}_{1}^{2}-q_{{2}}p_{{1}}p_{{2}}-q_{{3}}p_{{1}}p_{{3}}-q_{{3}}\frac{1}{2}\,{p}_{2}^{2}, \\ \mathcal{E}_{2}&=\frac{1}{2}\,q_{3}^{2}{p}_{3}^{2}+q_{{2}}q_{{3}}p_{{2}}p_{{3}}-\frac{1}{2}q_{{2}}{p}_{1}^{2}+\frac{1}{2}\left( \,{q}_{2}^{2}-\,q_{{1}}q_{{3}}\right) {p}_{2}^{2}-q_{{3}}p_{{1}}p_{{2}}, \\ \mathcal{E}_{3}&=-\frac{1}{2}\,q_{{3}}{p}_{1}^{2}+\frac{1}{2}q_{{2}}q_{{3}}\,{p}_{2}^{2}+{q}_{3}^{2}p_{{2}}p_{{3}}. \end{aligned} \label{geo8}%
\end{equation}

Finally, for $m=4$, we have $\mathfrak{a}=\mathfrak{g}$, so $H_{r}^{A}%
=h_{r}^{A}$ for all $r$ and
\begin{align}
h_{r}^{A}  &  =\mathcal{E}_{r}+a_{{4}}{\mathrm{e}^{2t_{{1}}}}V_{r}%
^{(4)}+(a_{{3}}{\mathrm{e}^{t_{{1}}}}+a_{4}{\mathrm{e}^{t_{{1}}}})V_{r}%
^{(3)}+[a_{-1}{+2a}_{-2}t_{2}+a_{-3}(t_{3}+3t_{2}^{2})\nonumber\\
&  \quad{} +4a_{-4}(t_{2}t_{3}+t_{2}^{3}){]V}_{r}^{(-1)}+[a_{-2}+3a_{-3}%
t_{2}+2a_{-4}(t_{3}+3t_{2}^{2})]{V}_{r}^{(-2)}\nonumber\\
&  \quad{} +(a_{-3}+4a_{-4}t_{2}){V}_{r}^{(-3)}+a_{-4}V_{r}^{(-4)}%
-c_{n-r}(t_{1},t_{2},t_{3},a_{0},a_{1},a_{2},a), \label{3.4}%
\end{align}
where
\begin{equation}
\begin{aligned} \mathcal{E}_{1}&=\frac{1}{2}\left( \,{q}_{1}^{2}-\,q_{{2}}\right) {p}_{1}^{2}+\frac{1}{2}\,{q}_{2}^{2}{p}_{2}^{2}+\frac{1}{2}\,q_{3}^{2}{p}_{3}^{2}+\left( q_{{2}}q_{{1}}-q_{{3}}\right) p_{{1}}p_{{2}}+q_{{1}}q_{{3}}p_{{1}}p_{{3}}+q_{{2}}q_{{3}}p_{{2}}p_{{3}}, \\ \mathcal{E}_{2}&=\frac{1}{2}\left( q_{{1}}\,q_{{2}}-\,q_{{3}}\right) {p}_{1}^{2}+{p}_{2}^{2}q_{{3}}q_{{2}}+q_{2}^{2}p_{{1}}p_{{2}}+q_{{2}}q_{{3}}p_{{1}}p_{{3}}+q_{3}^{2}p_{{2}}p_{{3}}+q_{2}p_{1}+2q_{3}p_{2}, \\ \mathcal{E}_{3}&=q_{2}q_{3}p_{{1}}p_{{2}}+q_{3}^{2}p_{{1}}p_{{3}}+\frac{1}{2}\,q_{{3}}q_{{1}}{p}_{1}^{2}+\frac{1}{2}\,q_{3}^{2}{p}_{2}^{2}+q_{3}p_{1}. \end{aligned} \label{geo9}%
\end{equation}
The metric $G$ is flat for $m=0,1,2,3$, so for these cases we can also express
all the formulas in flat coordinates $(x,y)$.

\begin{example}
\label{2e}Non-autonomous H\'{e}non-Heiles system. Consider the system
generated by Hamiltonians (\ref{e2}) with $a_{5}=a_{3}=a_{2}=0$,
$a_{-1}=-\frac{1}{4}\alpha$ and $a_{4}=-1$
\begin{align*}
h_{1}  &  =\frac{1}{2}\,{p}_{1}^{2}-\frac{1}{2}\,q_{{2}}{p}_{2}^{2}-q_{1}%
^{3}+2q_{1}q_{2}+3t_{2}(q_{1}^{2}-q_{2})-(t_{1}+3t_{2}^{2})q_{1}-\frac{1}%
{4}\alpha\,q_{2}^{-1}-c_{1}(t_{1},t_{2,}a_{0,}a_{1}),\\
h_{2}  &  =-q_{{2}}p_{{2}}p_{{1}}-\frac{1}{2}\,q_{{1}}q_{{2}}{p}_{2}^{2}%
+p_{1}+q_{2}^{2}-q_{1}^{2}q_{2}+3t_{2}q_{1}q_{2}-(t_{1}+3t_{2}^{2})q_{2}%
-\frac{1}{4}\alpha\,q_{1}q_{2}^{-1}-c_{0}(t_{1},t_{2,}a_{0,}a_{1}),
\end{align*}
where the functions $c_{r}$ are such that $h_{r}$ satisfy the Frobenius
condition (\ref{fc}). For example, a possible choice is%
\begin{equation}
a_{0}=0,\quad a_{1}=0,\quad c_{0}=\frac{1}{2}t_{1}^{2}+3t_{1}t_{2}^{2},\quad
c_{1}=0. \label{wybor}%
\end{equation}
Another choice is%
\[
a_{0}=a_{1}=1,\quad c_{1}=-t_{2}^{3},\quad c_{0}=\frac{1}{2}t_{1}^{2}.
\]
In flat orthogonal coordinates $(x,y)$
\[
q_{1}=-x_{1},\quad q_{2}=-\frac{1}{4}x_{2}^{2},\quad p_{1}=-y_{1},\quad
p_{2}=-\frac{2y_{2}}{x_{2}},
\]
the Hamiltonians $h_{i}$ take the form
\begin{align*}
h_{1}  &  =\frac{1}{2}y_{1}^{2}+\frac{1}{2}y_{2}^{2}+x_{1}^{3}+\frac{1}%
{2}x_{1}x_{2}^{2}+\alpha\,x_{2}^{-2}+3t_{2}(x_{1}^{2}+\frac{1}{4}x_{2}%
^{2})+(t_{1}+3t_{2}^{2})x_{1}-c_{1}\\
&  =h_{1}^{HH}+3t_{2}(x_{1}^{2}+\frac{1}{4}x_{2}^{2})+(t_{1}+3t_{2}^{2}%
)x_{1}-c_{1},\\
h_{2}  &  =\frac{1}{2}x_{2}y_{1}y_{2}-\frac{1}{2}x_{1}y_{2}^{2}-y_{1}+\frac
{1}{16}x_{2}^{4}+\frac{1}{4}x_{1}^{2}x_{2}^{2}-\alpha\,x_{1}x_{2}^{-2}%
+\frac{1}{4}3t_{2}x_{1}x_{2}^{2}+\frac{1}{4}(t_{1}+3t_{2}^{2})x_{2}^{2}%
-c_{0}\\
&  =h_{2}^{HH}+\frac{1}{4}3t_{2}x_{1}x_{2}^{2}+\frac{1}{4}(t_{1}+3t_{2}%
^{2})x_{2}^{2}-c_{0}%
\end{align*}
and constitute a non-autonomous deformation of the integrable case of the
extended H\'{e}non-Heiles system $h_{r}^{HH}$. Moreover, the flow generated by
$h_{1}^{HH}$ is exactly the stationary flow of the $5th$-order KdV
\cite{Allan}.
\end{example}

\section{Frobenius integrable deformations of quasi-St\"{a}ckel systems with
magnetic potentials\label{sec6}}

In this section we are going to construct, in a systematic way,
multi-parameter families of Frobenius integrable (satisfying (\ref{fc}))
non-autonomous Hamiltonian systems with the so called magnetic potentials
(vector potentials). Consider the following quasi-separation relations
\begin{equation}
\sum_{\gamma\in B}d_{\gamma}(t_{1},\dotsc,t_{n})\lambda_{i}^{\gamma}\mu
_{i}+\sum_{r=1}^{n}\lambda_{i}^{n-r}h_{r}^{B}=\frac{1}{2}\lambda_{i}^{m}%
\mu_{i}^{2}+\sum_{k=1}^{n}v_{ik}(\lambda)\mu_{k},\quad i=1,\ldots,n,\quad
B\subset\mathbb{Z}, \label{sepm}%
\end{equation}
(compare with (\ref{sepg})) where as before $m\in\{0,\dots,n+1\}$ while
$v_{ik}$ are given by (\ref{ni}). The system (\ref{sepm}) again naturally
generalizes (\ref{sep}). Solving (\ref{sepm}) with respect to $h_{r}^{B}$\ we
obtain $n$ Hamiltonians
\begin{equation}
h_{r}^{B}=E_{r}+W_{r}+M_{r}^{B}=\mathcal{E}_{r}+M_{r}^{B},\quad r=1,\ldots,n
\label{hn}%
\end{equation}
where $E_{r}$, $W_{r}$ are given respectively by (\ref{4}) and (\ref{kil5}).
The functions%
\[
M_{r}^{B}=\sum_{\gamma\in B}d_{\gamma}(t_{1},\dotsc,t_{n})M_{r}^{(\gamma)}%
\]
are time-dependent linear combinations of what we will call \emph{basic
magnetic separable potentials} $M_{r}^{(\gamma)}$ (see below). By linearity of
(\ref{sepm}), the potentials $M_{r}^{(\gamma)}$ satisfy the relations%
\[
\lambda_{i}^{\gamma}\mu_{i}+\sum_{r=1}^{n}\lambda_{i}^{n-r}M_{r}^{(\gamma
)}=0,\quad i=1,\ldots,n
\]
and by Lemma \ref{VDM} they have the linear in momenta $\mu_{i}$ form
\begin{equation}
M_{r}^{(\gamma)}=\sum_{i=1}^{n}\frac{\partial\rho_{r}}{\partial\lambda_{i}%
}\frac{\lambda_{i}^{\gamma}\mu_{i}}{\Delta_{i}}=\sum_{i=1}^{n}(P_{r}%
^{(\gamma)})^{i}\mu_{i}=\mu^{T}P_{r}^{(\gamma)}, \label{1b}%
\end{equation}
where $P_{r}^{(\gamma)}$ are vector fields (vector potentials) on the base
manifold $Q$ given by%
\begin{equation}
P_{r}^{(\gamma)}=\sum_{i=1}^{n}\frac{\partial\rho_{r}}{\partial\lambda_{i}%
}\frac{\lambda_{i}^{\gamma}}{\Delta_{i}}\frac{\partial}{\partial\lambda_{i}}.
\end{equation}
and so $\mu^{T}P_{r}^{(\gamma)}$ are linear in momenta functions on
$\mathcal{M}=T^{\ast}Q$ induced by the corresponding vector fields
$P_{r}^{(\gamma)}$.

\begin{remark}
\label{odpr}We call the vector fields $P_{r}^{(\gamma)}$ (vector) magnetic
potentials as they can be obtained by shifting momenta in non-magnetic
Hamiltonians (\ref{hg1}), as it is customary in literature. By a slight abuse
of terminology we then also call the functions $M_{r}^{(\gamma)}$ on
$\mathcal{M}$ scalar magnetic potentials. The functions $M_{r}^{(\gamma)}$ are
linear in momenta and separable terms in Hamiltonians. Note also that the
quasi-St\"{a}ckel terms $W_{r}=\mu^{T}J_{r}$ in (\ref{sepm}) are also linear
in momenta, but they are not separable and moreover they play a distinguished
role in our construction, so we do not call them magnetic.
\end{remark}

Using the relations
\begin{equation}
\frac{\partial\rho_{r}}{\partial\lambda_{i}}=-\sum_{s=0}^{r-1}\rho_{s}%
\lambda_{i}^{r-s-1},\qquad\sum_{i=1}^{n}\frac{\lambda_{i}^{k}\mu_{i}}%
{\Delta_{i}}=\sum_{j=1}^{n}V_{j}^{(k)}p_{j}, \label{potrzebne}%
\end{equation}
we can immediately rewrite (\ref{1b}) in Vi\`{e}te coordinates (\ref{26}) as
\[
M_{r}^{(\gamma)}=p^{T}P_{r}^{(\gamma)}=-\sum_{j=1}^{n}\left[  \sum_{s=0}%
^{r-1}q_{s}V_{j}^{(r+\gamma-s-1)}\right]  p_{j}%
\]
(note that $p^{T}P_{r}^{(\gamma)}$ is the same function on $\mathcal{M}$ as
$\mu^{T}P_{r}^{(\gamma)}$, only written in Vi\`{e}te coordinates). We will now
formulate the following problem: (cf. Problem \ref{problem}).

\begin{problem}
\label{problem2}For an arbitrary $m\in\{0,\ldots,n+1\}$, determine the set
$B$, as well as the explicit form of the coefficients $d_{\gamma}(t_{1}%
,\dotsc,t_{n})$ such that the system (\ref{hn}) can be deformed by formulas
(\ref{7a})--(\ref{7b}) to a Frobenius integrable non-autonomous system
satisfying (\ref{fc}).
\end{problem}

In order to solve this problem let us first consider Hamiltonians (\ref{hn})
with a single basic magnetic potential $M_{r}^{(\gamma)}$ with an arbitrary
$\gamma\in\mathbb{Z}$:%
\begin{equation}
h_{r}^{(\gamma)}=\mathcal{E}_{r}+M_{r}^{(\gamma)}, \quad r=1,\ldots,n
\label{jola1}%
\end{equation}
(which corresponds to choosing the monomial $\lambda_{i}^{\gamma}\mu_{i}$ in
the left hand side of (\ref{sepm})). It can be proved that $h_{r}^{(\gamma)}$
satisfy a magnetic analogue of Theorem \ref{1t}:

\begin{theorem}
\label{2t} Denote%
\[
a_{\gamma,k}=%
\begin{cases}
\max\left\{  0,\gamma+k-n-1\right\}  & \text{if $k\in\left\{  1\right\}  \cup
I_{1}^{m}$}\\
\min\left\{  0,\gamma+k-n-1\right\}  & \text{if $k\in I_{2}^{m}$}.
\end{cases}
\]
For any $\gamma\in\left\{  0,\dotsc,n+1\right\}  $ the following commutation
relations (Poisson brackets) between the Hamiltonians (\ref{jola1}) hold:

\begin{description}
\item[(i)] for $r,s\in\left\{  1\right\}  \cup I_{1}^{m}$
\begin{equation}
\left\{  h_{r}^{(\gamma)},h_{s}^{(\gamma)}\right\}  =(s-r)h_{r+s+m-n-2}%
^{(\gamma)}-a_{\gamma,s}M_{r}^{(\gamma+s+m-n-2)}+a_{\gamma,r}M_{s}%
^{(\gamma+r+m-n-2)} \label{m1}%
\end{equation}

\item[(ii)] for $r,s\in I_{2}^{m}$
\begin{equation}
\left\{  h_{r}^{(\gamma)},h_{s}^{(\gamma)}\right\}  =-(s-r)h_{r+s+m-n-2}%
^{(\gamma)}+a_{\gamma,s}M_{r}^{(\gamma+s+m-n-2)}-a_{\gamma,r}M_{s}%
^{(\gamma+r+m-n-2)} \label{m2}%
\end{equation}

\item[(iii)] for $r\in\left\{  1\right\}  \cup I_{1}^{m}$, $s\in I_{2}^{m}$
\begin{equation}
\left\{  h_{r}^{(\gamma)},h_{s}^{(\gamma)}\right\}  =%
\begin{cases}
a_{\gamma,s}M_{r}^{(\gamma+s+m-n-2)} & \text{if $a_{\gamma,s},a_{\gamma,r}<0$%
}\\
a_{\gamma,r}M_{s}^{(\gamma+r+m-n-2)} & \text{if $a_{\gamma,s},a_{\gamma,r}>0$%
}\\
0 & \text{otherwise}%
\end{cases}
\label{m3}%
\end{equation}

\end{description}
\end{theorem}

It can be verified that for other combinations of indices $r,s,m$ and $\gamma$
there are no formulas allowing to express the commutator of two Hamiltonians
$h_{m,r}^{(\gamma)}$ as a linear combination of a Hamiltonian and basic
separable magnetic potentials.

One can prove this theorem analogously to the proof of Theorem \ref{1t}, i.e.
by direct calculation, but the validity of this theorem follows also from the
relations, described in Section \ref{sec8}, between systems with magnetic
potentials considered in the present section and the systems with ordinary
potentials, considered in Section~\ref{sec4}.

As we see from the above theorem, for an arbitrary $\gamma\in\{0,\dotsc,n+1\}$
the additional magnetic potentials $M_{r}^{(\delta)}$ on the right hand sides
of (\ref{m1})--(\ref{m3}) are such that $\delta$ always belongs to
$\{0,\dotsc,n+1\}$. This leads to the conclusion, that no matter what
$m\in\{0,\dots,n+1\}$ is, the set $B$ must be of the form%
\[
B=B_{\gamma}=\left\{  0,\ldots,\gamma\right\}  , \quad\gamma\leq n+1\text{ \ }%
\]
so that the maximal possible set $B$ is $B=\left\{  0,\ldots,n+1\right\}  $.
Having chosen $\gamma\leq n+1$ we obtain $h_{r}^{B}$ in (\ref{hn}) of the form%
\[
h_{r}^{B}=\mathcal{E}_{r}+M_{r}^{B}=E_{r}+W_{r}+\sum_{k=0}^{\gamma}d_{k}%
(t_{1},\dotsc,t_{n})M_{r}^{k},\quad r=1,\ldots,n
\]
where $B=\{0,\dotsc,\gamma\}$. The functions $d_{k}(t)$ can now be determined
from our four-step procedure, presented in Section \ref{sec4}. This procedure
will in the magnetic case yield a compatible and determined (contrary to the
non-magnetic case) system of PDE's for $d_{k}$.

\begin{example}
\label{3e}Let us illustrate this procedure in the case $n=3$, $m=1$ and
$\gamma=3$. Then the magnetic quasi-St\"{a}ckel Hamiltonians $h_{r}^{B}$ in
(\ref{hn}) attain the form%
\[
h_{r}^{B}=\mathcal{E}_{r}+\sum_{k=0}^{3}d_{k}(t_{1},\dotsc,t_{n})M_{r}^{(k)}.
\]
where in Vi\`{e}te coordinates $\mathcal{E}_{r}$ are exactly as in the
non-magnetic case, i.e. given by (\ref{E1a}) and where $M_{r}^{(k)}$ are given
by (\ref{1b}) and (\ref{2b}). In the first three steps of our procedure we
deform the Hamiltonians $h_{r}^{B}$ to the Hamiltonians $H_{r}^{B}$ given by
(\ref{E3}). In the last step we find the functions $d_{k}$ from the Frobenius
condition (\ref{fc}). Inserting $H_{r}^{B}$ given by (\ref{E3}) into the left
hand side of (\ref{fc}) and demanding that the results do not depend on the
phase space coordinates we obtain a compatible set of first order PDE's on
$d_{0},\dotsc,d_{3}$ that can be again solved recursively. The solution is as
follows
\[
d_{0}(t)=0,\quad d_{1}(t)=b_{3}(t_{2}+t_{3}^{2}),\quad d_{2}(t)=2b_{3}%
t_{3},\quad d_{3}(t)=b_{3}\in\mathbb{R},
\]
where we choose the integration constants $b_{0}=b_{1}=b_{2}=0$.
\end{example}

In general, applying our procedure we obtain a $(n+2$)-parameter family of
Frobenius integrable systems, parametrized by the integration constants
$(b_{0},\dotsc,b_{n+1})$ (in Example \ref{3e} above we have $b_{0}=b_{1}%
=b_{2}=b_{4}=0$ and $b_{3}\in\mathbb{R}$). In the magnetic case the
Hamiltonians have no ``tails'' depending on some non-dynamical variables,
contrary to the non-magnetic case.

\section{Frobenius integrable deformations of two- and three-dimensional
quasi-St\"{a}ckel systems with magnetic potentials\label{sec7}}

In this section we present a complete list of Frobenius integrable
deformations (\ref{7b}),(\ref{7c}) of two- and three-dimensional
quasi-St\"{a}ckel systems with magnetic potentials that originate in our
deformation procedure. We always use the maximal set $B=\left\{
0,\ldots,n+1\right\}  $. In each case $(n,m)$ we obtain a $(n+3)$-parameter
family of systems, parametrized by real constants $b_{0},\ldots,b_{n+1}%
,\overline{b}$.

\subsection{Two-dimensional systems}

Let us first consider the case $n=2$ so that $B=\left\{  0,\ldots,3\right\}
$. As it was explained in Section \ref{bla} in this case $H_{r}^{B}=h_{r}^{B}$
for each $m=0,\dotsc,3$ and for each $m$ we obtain a $4$-parameter family of
Frobenius integrable non-autonomous Hamiltonian systems satisfying (\ref{fc}).

Explicitly, for $m=0$ in Vi\`{e}te coordinates we get
\begin{equation}
h_{r}^{B}=\mathcal{E}_{r}+b_{3}M_{r}^{(3)}+b_{2}M_{r}^{(2)}+(b_{1}+2b_{3}%
t_{2})M_{r}^{(1)}+(b_{0}+b_{2}t_{2}+b_{3}t_{1})M_{r}^{(0)},\nonumber
\end{equation}
where $\mathcal{E}_{r}$ are given by (\ref{geo1}).

For $m=1$%
\[
h_{r}^{B}=\mathcal{E}_{r}+b_{3}M_{r}^{(3)}+(b_{2}+2b_{3}t_{2})M_{r}%
^{(2)}+[b_{1}+b_{2}t_{2}+b_{3}(t_{1}+t_{2}^{2})]M_{r}^{(1)}+b_{0}M_{r}^{(0)},
\]
where $\mathcal{E}_{r}$ are given by (\ref{geo2}).

For $m=2$%
\[
h_{r}^{B}=\mathcal{E}_{r}+b_{3}M_{r}^{(3)}+(b_{2}+b_{3}t_{1})M_{r}^{(2)}%
+b_{1}M_{r}^{(1)}+b_{0}\exp(t_{2})M_{r}^{(0)},
\]
where $\mathcal{E}_{r}$ are given by (\ref{geo3}).

For $m=3$%
\[
h_{r}^{B}=\mathcal{E}_{r}+M_{r}^{(4)}+b_{3}\exp(t_{1})M_{r}^{(3)}+b_{2}%
M_{r}^{(2)}+(b_{1}+b_{0}t_{2})M_{r}^{(1)}+b_{0}M_{r}^{(0)},
\]
where $\mathcal{E}_{r}$ are given by (\ref{geo4}). Moreover, in Vi\`{e}te
coordinates,%
\begin{align*}
M_{1}^{(0)}  &  =p_{2}, & M_{2}^{(0)}  &  =p_{1}+q_{1}p_{2},\\
M_{1}^{(1)}  &  =p_{1}, & M_{2}^{(1)}  &  =-q_{2}p_{2},\\
M_{1}^{(2)}  &  =-q_{1}p_{1}-q_{2}p_{2}, & M_{2}^{(2)}  &  =-q_{2}p_{1},\\
M_{1}^{(3)}  &  =(q_{1}^{2}-q_{2})p_{1}+q_{1}q_{2}p_{2}, & M_{2}^{(3)}  &
=q_{1}q_{2}p_{1}+q_{2}^{2}p_{2}.
\end{align*}
The metric $G$ is flat for $m=0,1,2$, so for these cases we can express all
the formulas in the flat coordinates $(x,y)$.

\subsection{Three-dimensional systems}

Let us now consider the case $n=3$ so that $B=\left\{  0,\ldots,4\right\}  $.
For each $m=0,\dotsc,4$ we obtain a $5$-parameter family of Frobenius
integrable non-autonomous Hamiltonian systems satisfying (\ref{fc}). For $m=0$
we have $\mathfrak{a}$ $=\mathfrak{g}$ so $H_{r}^{B}=h_{r}^{B}$ for all $r$
and our procedure yields
\begin{align}
h_{r}^{B}  &  =\mathcal{E}_{r}+b_{4}M_{r}^{(4)}+b_{3}M_{r}^{(3)}+(b_{2}%
+3b_{4}t_{3})M_{r}^{(2)}+(b_{1}+2b_{3}t_{3}+2b_{4}t_{2})M_{r}^{(1)}\nonumber\\
&  \quad{}+[b_{0}+b_{2}t_{3}+b_{3}t_{2}+b_{4}(\frac{3}{2}t_{3}^{2}%
+t_{1})]M_{r}^{(0)}, \label{v0}%
\end{align}
where $\mathcal{E}_{r}$ are given by (\ref{geo5}).

For $m=1$, we have, by (\ref{31}), $H_{1}^{B}=h_{1}^{B},~H_{2}^{B}=h_{2}^{B}$
and $H_{3}^{B}=h_{3}^{B}+t_{2}h_{1}^{B}$ where%
\begin{align}
h_{r}^{B}  &  =\mathcal{E}_{r}+b_{4}M_{r}^{(4)}+(b_{3}+3b_{4}t_{3})M_{r}%
^{(3)}+[b_{2}+2b_{3}t_{3}+b_{4}(3t_{3}^{2}+2t_{2})]M_{r}^{(2)}\nonumber\\
&  \quad{} +[b_{1}+b_{2}t_{3}+b_{3}(t_{3}^{2}+t_{2})+b_{4}(t_{3}^{3}%
+3t_{2}t_{3}+t_{1})]M_{r}^{(1)}+b_{0}M_{r}^{(0)}, \label{v1}%
\end{align}
where $\mathcal{E}_{r}$ are given by (\ref{geo6}).

For $m=2$ we have $\mathfrak{a}$ $=\mathfrak{g}$ so $H_{r}^{B}=h_{r}^{B}$ for
all $r$ with%
\begin{align*}
h_{r}^{B}  &  =\mathcal{E}_{r}+b_{4}M_{r}^{(4)}+(b_{3}+2b_{4}t_{2})M_{r}%
^{(3)}+[b_{2}+b_{3}t_{2}+b_{4}(t_{2}^{2}+t_{1})]M_{r}^{(2)}\\
&  \quad{} +b_{1}M_{r}^{(1)}+b_{0}\exp(t_{3})M_{r}^{(0)},
\end{align*}
where $\mathcal{E}_{r}$ are given by (\ref{geo7}).

For $m=3$ we have, due to (\ref{33}), $H_{1}^{B}=h_{1}^{B},~H_{3}^{B}%
=h_{3}^{B}$ and $H_{2}^{B}=h_{2}^{B}+t_{3}h_{3}^{B}$ with%
\begin{align*}
h_{r}^{B}  &  =\mathcal{E}_{r}+b_{4}M_{r}^{(4)}+(b_{3}+b_{4}t_{1})M_{r}%
^{(3)}+b_{2}M_{r}^{(2)}\\
&  \quad{} +[b_{0}t_{3}\exp(2t_{2})+b_{1}\exp(t_{2})]M_{r}^{(1)}+b_{0}%
\exp(2t_{2})M_{r}^{(0)},
\end{align*}
where $\mathcal{E}_{r}$ are given by (\ref{geo8}).

Finally, for $m=4$ we have $\mathfrak{a}=\mathfrak{g}$ so $H_{r}^{B}=h_{r}%
^{B}$ for all $r$ with%
\begin{align}
h_{r}^{B}  &  =\mathcal{E}_{r}+b_{4}\exp(t_{1})M_{r}^{(4)}+b_{3}M_{r}%
^{(3)}+[b_{2}+b_{1}t_{2}+b_{0}(t_{2}^{2}+t_{3})]M_{r}^{(2)}\nonumber\\
&  \quad{} +(b_{1}+2b_{0}t_{2})M_{r}^{(1)}+b_{0}M_{r}^{(0)}, \label{v4}%
\end{align}
where $\mathcal{E}_{r}$ are given by (\ref{geo9}).

Moreover, in Vi\`{e}te coordinates,%
\begin{equation}
\begin{aligned} M_{1}^{(0)}&=p_{3}, & M_{2}^{(0)}&=p_{2}+q_{1}p_{3}, & M_{3}^{(0)}&=p_{1}+q_{1}p_{2}+q_{2}p_{3}, \\ M_{1}^{(1)}&=p_{2}, & M_{2}^{(1)}&=p_{1}+q_{1}p_{2}, & M_{3}^{(1)}&=-q_{3}p_{3}, \\ M_{1}^{(2)}&=p_{1}, & M_{2}^{(2)}&=-q_{2}p_{2}-q_{3}p_{3}, & M_{3}^{(2)}&=-q_{3}p_{2}, \\ M_{1}^{(3)}&=-q_{1}p_{1}-q_{2}p_{2}-q_{3}p_{3}, & M_{2}^{(3)}&=-q_{2}p_{1}-q_{3}p_{2}, & M_{3}^{(3)}&=-q_{3}p_{1}, \\ M_{1}^{(4)}&=(q_{1}^{2}-q_{2})p_{1}+(q_{1}q_{2}-q_{3})p_{2} & M_{2}^{(4)}&=(q_{1}q_{2}-q_{3})p_{1}+q_{2}^{2}p_{2} & M_{3}^{(4)}&=q_{1}q_{3}p_{1}+q_{2}q_{3}p_{2} \\ & \quad {} +q_{1}q_{3}p_{3}, & & \quad {} +q_{2}q_{3}p_{3}, & & \quad {} +q_{3}^{2}p_{3}. \end{aligned} \label{2b}%
\end{equation}

As previously, the metric $G$ is flat for $m=0,1,2,3$, so for these cases we
can express all the formulas in the flat coordinates $(x,y)$.

\section{Canonical transformations between both classes of non-au\-tonomous
Frobenius integrable systems\label{sec8}}

In this section we construct \emph{multitime-dependent} canonical
transformations (rational symplectic transformations \cite{Iwasaki}) between
Frobenius integrable systems constructed in Section \ref{sec4} and Frobenius
integrable systems constructed in Section \ref{sec6}. This will allow us to
find the set of first order PDE's, described in our four-step procedure, for
the functions $d_{\gamma}(t_{1},\dotsc,t_{n})$, $\zeta_{r,j}(t_{1}%
,\dotsc,t_{r-1})$ and $\zeta_{r,r+j}(t_{r+1},\dotsc,t_{n})$ \emph{in an
explicit form}.

In the case of the ordinary potentials, considered in Section \ref{sec4}, we
have constructed a $(2n+3)$-parameter family of non-autonomous Frobenius
integrable systems with $n+3$ dynamical parameters (and $n$ non-dynamical
parameters) and in the case of systems with magnetic potentials, considered in
Section \ref{sec6}, we obtained a $(n+2)$-parameter family (all parameters are
dynamical). In order to relate both classes through a canonical transformation
we have thus to extend the systems with magnetic potentials by an additional
parameter, to make equal the number of dynamical parameters. Instead of
(\ref{sepm}), we consider thus the following quasi-separation relations,
obtained by extending (\ref{sepm}) by the simplest nontrivial ordinary
separable potential generated by $e(t_{1},\dotsc,t_{n})\lambda_{i}^{n}$:
\begin{equation}
e(t_{1},\dotsc,t_{n})\lambda_{i}^{n}+\sum_{\gamma=0}^{n+1}d_{\gamma}%
(t_{1},\dotsc,t_{n})\lambda_{i}^{\gamma}\mu_{i}+\sum_{r=1}^{n}\lambda
_{i}^{n-r}h_{r}^{B}=\frac{1}{2}\lambda_{i}^{m}\mu_{i}^{2}+\sum_{k=1}^{n}%
v_{ik}(\lambda)\mu_{k},\quad i=1,\ldots,n, \label{7.1}%
\end{equation}
so that, as before, $B=\{0,\dotsc,n+1\}$ but now
\begin{equation}
h_{r}^{B}=E_{r}+W_{r}+\sum_{\gamma=0}^{n+1}d_{\gamma}(t_{1},\dotsc,t_{n}%
)M_{r}^{(\gamma)}+e(t_{1},\dotsc,t_{n})V_{r}^{(n)},\quad r=1,\ldots,n.
\label{hm1}%
\end{equation}
Let us first discuss the dependence of $e(t_{1},\dotsc,t_{n})$ on times
$t_{r}$. From Theorem \ref{1t} and the Frobenius condition (\ref{fc}) it
follows that for the time-dependent coefficient $e$ at $\lambda_{i}^{n}$ in
the non-magnetic system
\begin{equation}
\frac{\partial e(\lambda)}{\partial t_{r}}=0\text{ \ for $m=0,\dotsc,n$%
},\qquad\frac{\partial e(\lambda)}{\partial t_{r}}=e(\lambda)\delta
_{1,r}\text{ \ for $m=n+1.$} \label{27d}%
\end{equation}
Thus, $e(t_{1},\dotsc,t_{n})=\overline{b}=\text{const}$ for $m=0,\dotsc,n$ and
$e(t_{1},\dotsc,t_{n})=\overline{b}\exp(t_{1})$ for $m=n+1$.

Rearranging terms in (\ref{7.1}), we obtain
\begin{equation}
\sum_{\alpha=-m}^{2n-m+2}c_{\alpha}^{\prime}(t)\lambda_{i}^{\alpha}+\sum
_{r=1}^{n}\lambda_{i}^{n-r}h_{r}^{B}=\frac{1}{2}\lambda_{i}^{m}\left(  \mu
_{i}-\sum_{\gamma=0}^{n+1}d_{\gamma}(t)\lambda_{i}^{\gamma-m}\right)
^{2}+\sum_{k=1}^{n}v_{ik}(\lambda)\mu_{k},\quad i=1,\ldots,n, \label{7.1a}%
\end{equation}
where $2n+3$ functions $c_{\alpha}^{\prime}(t)$ are uniquely defined through
the polynomial relation
\begin{equation}
\sum_{\alpha=-m}^{2n-m+2}c_{\alpha}^{\prime}(t)\lambda^{\alpha}=\frac{1}%
{2}\lambda^{m}\left(  \sum_{\gamma=0}^{n+1}d_{\gamma}(t)\lambda^{\gamma
-m}\right)  ^{2}+e(t)\lambda^{n}. \label{7.10}%
\end{equation}
The form (\ref{7.1a}) of quasi-separation relations (\ref{7.1}) indicates the
existence of a very natural, time dependent change of coordinates, that will
allow us to rewrite (\ref{7.1a}) as a \emph{non-magnetic} system from the
class (\ref{sepg}). Indeed, let us rewrite quasi-separation relations
(\ref{7.1a}) in new variables $\lambda_{i}^{\prime}$, $\mu_{i}^{\prime}$
related to $\lambda_{i}$, $\mu_{i}$ through the canonical transformation
depending on all times
\begin{equation}
\lambda_{i}^{\prime}=\frac{\partial F(\lambda,\mu^{\prime},t)}{\partial\mu
_{i}^{\prime}}=\lambda_{i},\quad\mu_{i}=\frac{\partial F(\lambda,\mu^{\prime
},t)}{\partial\lambda_{i}}=\mu_{i}^{\prime}+\sum_{\gamma=0}^{n+1}d_{\gamma
}(t_{1},\dotsc,t_{n})\lambda_{i}^{\gamma-m},\quad i=1,\dotsc,n, \label{7.3}%
\end{equation}
generated by
\begin{equation}
F(\lambda,\mu^{\prime},t)=\sum_{i=1}^{n}\left(  \lambda_{i}\mu_{i}^{\prime
}+\sum_{\gamma=0,\gamma\neq m-1}^{n+1}\frac{1}{\gamma-m+1}d_{\gamma}%
(t)\lambda_{i}^{\gamma-m+1}+d_{m-1}(t)\ln\lambda_{i}\right)  , \label{7.4}%
\end{equation}
(we stress that this transformation depends on $m$). We find that Hamiltonians
$h_{r}^{B}$, calculated from new separation relations, are of the form%
\begin{equation}
h_{r}^{B}(\lambda,\mu^{\prime},t)=E_{r}^{\prime}+W_{r}^{\prime}+\sum
_{\alpha=-m}^{2n-m+2}c_{\alpha}^{\prime}(t)V_{r}^{(\alpha)}+S_{r}(t,\lambda),
\label{7.7}%
\end{equation}
where $E_{r}^{\prime}$ and $W_{r}^{\prime}$ are obtained from $E_{r}$, $W_{r}$
by replacing each $\mu_{i}$ by $\mu_{i}^{\prime}$, and where $S_{r}%
=W_{r}-W_{r}^{\prime}$. Note that (\ref{7.7}) do indeed have the non-magnetic
form (\ref{hg1}).

Let us now calculate the functions $S_{r}(t,\lambda)$ explicitly. In order to
do it we introduce the following notation:
\[
Z_{k}=\sum_{i=1}^{n}\lambda_{i}^{k},\quad k\in\mathbb{Z}%
\]
and further%
\begin{equation}
\begin{aligned} \mathcal{Z}_{1} &= d_{n+1}(t)Z_{1} & \\ \mathcal{Z}_{r} &= \sum_{k=1}^{r}d_{n-r+1+k}(t)Z_{k}, & r\in I_{1}^{m}, \\ \mathcal{Z}_{r} &= -\sum_{k=1}^{n-r+1}d_{n-r+1-k}(t)Z_{-k}, & r\in I_{2}^{m}. \end{aligned} \label{Zety}%
\end{equation}
Note that contrary to $Z_{k}$ the functions $\mathcal{Z}_{r}$ on $Q$ do depend
both on $t$ and on $m$. We are now in position to formulate the following theorem.

\begin{theorem}
\label{10t} The functions $S_{r}=W_{r}-W_{r}^{\prime}$ are given by
\begin{subequations}
\label{100a}%
\begin{align}
S_{1}  &  =-\mathcal{Z}_{1}-d_{n+1}(t)V_{1}^{(n)}=0, &  & \label{100aa}\\
S_{r}  &  =-(r-1)d_{n-r+1}(t)-\mathcal{Z}_{r}-d_{n+1}(t)V_{r}^{(n)}, & r  &
\in I_{1}^{m},\label{100b}\\
S_{r}  &  =(n-r+1)d_{n-r+1}(t)-\mathcal{Z}_{r}-d_{n+1}(t)V_{r}^{(n)}, & r  &
\in I_{2}^{m}, \label{100c}%
\end{align}

\end{subequations}
\end{theorem}

The proof of this theorem can be found in Appendix~B.

We demand that the Frobenius integrable systems, defined by the deformations
(\ref{7b}) and (\ref{7c}) of Hamiltonians (\ref{hg1}) and (\ref{hm1}), are
related by the multitime-dependent canonical transformation (\ref{7.3}). Thus,
according to the Hamilton-Jacobi theory of time-dependent canonical
transformations (rational symplectic transformations \cite{Iwasaki}), we have
for the same $m$
\begin{equation}
H_{r}^{A}(\lambda,\mu^{\prime},t)=H_{r}^{B}(\lambda,\mu^{\prime}%
,t)+\frac{\partial F(\lambda,\mu^{\prime},t)}{\partial t_{r}}, \label{7.5}%
\end{equation}
where, due to (\ref{7.4}),
\begin{equation}
\frac{\partial F(\lambda,\mu^{\prime},t)}{\partial t_{r}}=\sum_{\gamma
=0}^{n+1}\frac{1}{\gamma-m+1}\frac{\partial d_{\gamma}(t)}{\partial t_{r}%
}Z_{\gamma-m+1}. \label{7.6}%
\end{equation}
(there is no singularity in (\ref{7.6}) as $d_{m-1}$ is always constant).
Consider first the Hamiltonians $h_{r}^{B}$ for which $H_{r}^{B}=h_{r}^{B}$
(i.e. when $r\in\left\{  1,\ldots,\kappa_{1}\right\}  \cup\left\{
n-\kappa_{2}+1,\ldots,n\right\}  $, see again (\ref{7b}) and (\ref{7c})). In
such cases, according to (\ref{7.7}), (\ref{7.5}) and (\ref{jeden}), relations
(\ref{7.5}) boil down to
\begin{equation}
\sum_{\alpha=-m}^{2n-m+2}c_{\alpha}(t)V_{r}^{(\alpha)}=\sum_{\alpha
=-m}^{2n-m+2}c_{\alpha}^{\prime}(t)V_{r}^{(\alpha)}+S_{r}(t,\lambda
)+\frac{\partial F(\lambda,\mu^{\prime},t)}{\partial t_{r}},\quad
r=1,\dotsc,n, \label{beznumeru}%
\end{equation}
where $c_{\alpha}(t)$\ are coefficients of Hamiltonians $h_{r}^{A}$
(\ref{jeden}). In order to satisfy this demand, it is necessary and sufficient
to demand (since the terms in (\ref{beznumeru}) that contain basic separable
potentials must cancel on their own) both
\begin{equation}
c_{\alpha}(t)=c_{\alpha}^{\prime}(t),\ \ \ \ \ \ \alpha=-m,...,2n-m+2
\label{7.9a}%
\end{equation}
and
\begin{equation}
S_{r}(t,\lambda)+\frac{\partial F(\lambda,\mu^{\prime},t)}{\partial t_{r}}=0
\label{7.9aa}%
\end{equation}
separately. Now, from (\ref{7.9a}) we get a unique relation between the set of
coefficients $\{c_{-m}(t),...,c_{2n-m+2}(t)\}$ of Hamiltonians (\ref{jeden})
end the set of coefficients $\{d_{0}(t),...,d_{n+1}(t),e(t)\}$ of Hamiltonians
(\ref{hm1}) by simple comparing the Laurent polynomials in $\lambda$ from both
sides of (\ref{7.10}). Notice that in the case of canonical transformation the
set of non-dynamical coefficients $\{c_{0}(t),...,c_{n-1}(t)\}$ is uniquely
determined. The remaining conditions (\ref{7.9aa}) lead to a set of linear
first-order PDE's for functions $d_{\gamma}(t)$, $\zeta_{r,j}(t)$ and
$\zeta_{r,r+j}(t)$, respectively. Actually, according to (\ref{100aa}%
)--(\ref{100c}) and (\ref{Zety})%
\begin{equation}
\frac{\partial F}{\partial t_{r}}=\mathcal{Z}_{r},\quad r\in\{1,\dotsc
,\kappa_{1}\}\cup\{n+1-\kappa_{2},\dotsc,n\}. \label{7.9}%
\end{equation}
For the remaining values of $r$ we have to take in (\ref{7.9}) an appropriate
time dependent linear combinations of $\mathcal{Z}_{r}$\ derived for geodesic
parts in Section \ref{bla} (given by (\ref{7b}) and (\ref{7c})). Thus, we have
to require that
\begin{equation}
\begin{aligned} \frac{\partial F}{\partial t_{r}}&=\sum_{j=1}^{r}\zeta _{r,j}(t_{1},\dotsc,t_{r-1})\mathcal{Z}_{j}, & r\in \{\kappa _{1}+1,\dotsc,n-m+1\}\subset I_{1}^{m}, \\ \frac{\partial F}{\partial t_{r}}&=\sum_{j=0}^{n-r}\zeta _{r,r+j}(t_{r+1},\dotsc,t_{n})\mathcal{Z}_{n+j}, & r\in \{n-m+2,\dotsc,n-\kappa _{2}\}\subset I_{2}^{m}. \end{aligned} \label{7.12}%
\end{equation}
Comparing coefficients at $Z_{k}$ in (\ref{7.9}) and in (\ref{7.12}) on one
side and in (\ref{7.6}) on the other side we obtain, after some calculations,
the following theorem.

\begin{theorem}
\label{PDE}The Frobenius integrable systems defined by the deformations
(\ref{7b}) and (\ref{7c}) of Hamiltonians defined by (\ref{hg1}) and by
(\ref{hm1}) are related by the canonical transformation (\ref{7.3}) provided
that functions $d_{\gamma}$ satisfy the following set of linear first-order
PDE's: \newline1. For $r\in\{1,\dotsc,\kappa_{1}\}\subset\{1\}\cup I_{1}^{m}$
\begin{equation}
\begin{aligned} \frac{\partial d_{\gamma }}{\partial t_{r}}& =0, \quad \gamma \neq m,\dotsc,m+r-1 \\ \frac{\partial d_{\gamma }}{\partial t_{r}}& =(\gamma -m+1)d_{n-m+2+\gamma -r}, \quad \gamma =m,\dotsc,m+r-1. \end{aligned} \label{7.9b}%
\end{equation}
2. For $r\in\{\kappa_{1}+1,\dotsc,n-m+1\}\subset I_{1}^{m}$
\begin{equation}
\begin{aligned} \frac{\partial d_{\gamma }}{\partial t_{r}}& =0,\quad \gamma \neq m,\dotsc,m+r-1 \\ \frac{\partial d_{\gamma }}{\partial t_{r}}& =(\gamma -m+1)\sum_{j=\gamma -m+1}^{r}\zeta _{r,j}(t_{1},\dotsc,t_{r-1})d_{n-m+2+\gamma -j},\quad \gamma =m,\dotsc,m+r-1. \end{aligned} \label{7.12a}%
\end{equation}
3. For $r\in\{n-m+2,\dotsc,n-\kappa_{2}\}\subset I_{2}^{m}$
\begin{equation}
\begin{aligned} \frac{\partial d_{\gamma }}{\partial t_{r}}& =0,\quad \gamma \neq r-(n-m+2),\ldots ,m-2 \\ \frac{\partial d_{\gamma }}{\partial t_{r}}& =-(\gamma -m+1)\sum_{j=0}^{n-m+2+\gamma -r}\zeta _{r,r+j}(t_{r+1},\dotsc,t_{n})d_{n-m+2+\gamma -r-j},\quad \gamma =r-(n-m+2),\ldots ,m-2. \end{aligned} \label{7.12b}%
\end{equation}
4. For $r\in\{n+1-\kappa_{2},\dotsc,n\}\subset I_{2}^{m}$
\begin{equation}
\begin{aligned} \frac{\partial d_{\gamma }}{\partial t_{r}}& =0,\quad \gamma \neq r-(n-m+2),\ldots ,m-2 \\ \frac{\partial d_{\gamma }}{\partial t_{r}}& =-(\gamma -m+1)d_{n-m+2+\gamma -r},\quad \gamma =r-(n-m+2),\ldots ,m-2. \end{aligned} \label{7.9c}%
\end{equation}
Besides, the functions $\zeta_{r,j}(t_{1},\dotsc,t_{r-1})$ and $\zeta
_{r,r+j}(t_{r+1},\dotsc,t_{n})$ in (\ref{7b})--(\ref{7c}) can be calculated
from the system of first-order PDE's resulting from the compatibility
conditions
\begin{equation}
\frac{\partial^{2}d_{\gamma}}{\partial t_{r}\partial t_{s}}=\frac{\partial
^{2}d_{\gamma}}{\partial t_{s}\partial t_{r}},\quad r,s=1,\dotsc,n
\label{7.13}%
\end{equation}
if we choose all integration constants in (\ref{7.13}) to be zero.
\end{theorem}

The above system of first order PDE's for coefficients $\zeta_{r,j}$,
$\zeta_{r,r+j}$ and $d_{\gamma}$ can be solved recursively, giving explicit
time dependence for each $\zeta_{r,j}(t_{1},\dotsc,t_{r-1})$ and
$\zeta_{r,r+j}(t_{r+1},\dotsc,t_{n})$ from (\ref{7b}) and (\ref{7c}) and for
each $d_{\gamma}(t_{1},\dotsc,t_{n})$ from (\ref{7.1}). The same coefficients
were derived in Sections \ref{bla}, \ref{sec6} and \ref{sec7}, separately for
each case, directly from Frobenius conditions (\ref{fc}).

Comparising coefficients at equal powers of $x$ in (\ref{7.10}) we obtain the
map%
\begin{equation}
(b_{0},\ldots,b_{n+1},\overline{b})\rightarrow(a_{-m},\ldots,a_{-1}%
,a_{n},\ldots,a_{2n-m+2}) \label{mapka}%
\end{equation}
between the parameters $(b_{0},\ldots,b_{n+1},\overline{b})$ of the magnetic
representation and the dynamical parameters $(a_{-m},\ldots,a_{-1}%
,a_{n},\ldots,a_{2n-m+2})$ of the non-magnetic representation. It turns that
this map is not surjective so that not all non-magnetic systems have their
magnetic counterparts.

Finally, the coefficients $c_{\alpha}(t_{1},\dotsc,t_{n}),$ $\alpha
=-m,\dotsc,2n-m+2$ from (\ref{jeden}) are reconstructed by the relations
(\ref{mapka}), (\ref{7.9a}).

Let us illustrate this theorem by three examples.

\begin{example}
For $n=3$ and $m=0$ as $H_{r}^{B}=h_{r}^{B}$ for all $r$ so, according to
(\ref{7.9b}), we get the following set of first order PDE's for $d_{0}%
,\ldots,d_{4}$:%
\[%
\begin{array}
[c]{lll}%
\dfrac{\partial d_{4}}{\partial t_{k}}=0, & \ \dfrac{\partial d_{3}}{\partial
t_{k}}=0, & \ k=1,2,3,\\
& \  & \\
\dfrac{\partial d_{2}}{\partial t_{1}}=0, & \ \dfrac{\partial d_{2}}{\partial
t_{2}}=0, & \ \dfrac{\partial d_{2}}{\partial t_{3}}=3d_{4},\\
& \  & \\
\dfrac{\partial d_{1}}{\partial t_{1}}=0, & \ \dfrac{\partial d_{1}}{\partial
t_{2}}=2d_{4}, & \ \dfrac{\partial d_{1}}{\partial t_{3}}=2d_{3},\\
& \  & \\
\dfrac{\partial d_{0}}{\partial t_{1}}=d_{4}, & \ \dfrac{\partial d_{0}%
}{\partial t_{2}}=d_{3}, & \ \dfrac{\partial d_{0}}{\partial t_{3}}=d_{2},
\end{array}
\]
which are solved recursively to
\begin{gather*}
d_{4}(t)=b_{4},\quad d_{3}(t)=b_{3},\quad d_{2}(t)=b_{2}+3b_{4}t_{3},\quad
d_{1}(t)=b_{1}+2b_{3}t_{3}+2b_{4}t_{2},\\
d_{0}(t)=b_{0}+b_{2}t_{3}+b_{3}t_{2}+b_{4}(t_{1}+\tfrac{3}{2}t_{3}^{2})
\end{gather*}
obtained previously in (\ref{v0}). Moreover, using (\ref{7.10}) and
(\ref{7.9a}) we easily reconstruct the coefficients $c_{\alpha}(t_{1}%
,t_{2},t_{3})$ in Hamiltonians (\ref{3.0}), which yields the map (\ref{mapka})
that for dynamical parameters $a_{k}$ reads:
\begin{gather*}
a_{8}=\frac{1}{2}b_{4}^{2},\quad a_{7}=b_{3}b_{4},\quad a_{6}=\frac{1}{2}%
b_{3}^{2}+b_{2}b_{4},\quad a_{5}=b_{1}b_{4}+b_{2}b_{3},\\
a_{4}=\frac{1}{2}b_{2}^{2}+b_{0}b_{4}+b_{1}b_{3},\quad a_{3}=b_{0}b_{3}%
+b_{1}b_{2}+\overline{b}-b_{4}.
\end{gather*}

\end{example}

\begin{example}
For $n=3$ and $m=1$, as $H_{1}^{B}=h_{1}^{B},~H_{2}^{B}=h_{2}^{B}$ and
$H_{3}^{B}=h_{3}^{B}+\zeta_{1}(t_{1},t_{2})h_{1}^{B}+\zeta_{2}(t_{1}%
,t_{2})h_{2}^{B}$, so, according to (\ref{7.13}), (\ref{7.9b}) and
(\ref{7.12a}), we get the following set of first order PDE's%
\[%
\begin{array}
[c]{lll}%
\dfrac{\partial\zeta_{2}}{\partial t_{1}}=0, & \ \dfrac{\partial\zeta_{2}%
}{\partial t_{2}}=0, & \ \dfrac{\partial\zeta_{1}}{\partial t_{1}%
}=0,\ \ \ \ \ \ \ \dfrac{\partial\zeta_{1}}{\partial t_{2}}=1,\\
& \  & \\
\dfrac{\partial d_{4}}{\partial t_{k}}=0, & \ \dfrac{\partial d_{0}}{\partial
t_{k}}=0, & \ k=1,2,3,\\
& \  & \\
\dfrac{\partial d_{3}}{\partial t_{1}}=0, & \ \dfrac{\partial d_{3}}{\partial
t_{2}}=0, & \ \dfrac{\partial d_{3}}{\partial t_{3}}=3d_{4},\\
& \  & \\
\dfrac{\partial d_{2}}{\partial t_{1}}=0, & \ \dfrac{\partial d_{2}}{\partial
t_{2}}=2d_{4}, & \ \dfrac{\partial d_{2}}{\partial t_{3}}=2d_{3}+2\zeta
_{2}d_{4},\\
& \  & \\
\dfrac{\partial d_{1}}{\partial t_{1}}=d_{4}, & \ \dfrac{\partial d_{1}%
}{\partial t_{2}}=d_{3}, & \ \dfrac{\partial d_{1}}{\partial t_{3}}%
=d_{2}+\zeta_{2}d_{3}+\zeta_{1}d_{4},
\end{array}
\]
which are solved recursively to
\begin{equation}
\begin{gathered} \zeta _{2}(t)=0,\quad \zeta _{1}(t)=t_{2},\quad d_{4}(t)=b_{4},\quad d_{0}(t)=b_{0}, \\ d_{3}(t)=b_{3}+3b_{4}t_{3},\quad d_{2}(t)=b_{2}+2b_{3}t_{3}+b_{4}(3t_{3}^{2}+2t_{2}), \\ d_{1}(t)=b_{1}+b_{2}t_{3}+b_{3}(t_{3}^{2}+t_{2})+b_{4}(t_{3}^{3}+3t_{2}t_{3}+t_{1}), \end{gathered}\label{w1}%
\end{equation}
obtained previously in (\ref{v1}). Again, the formulas (\ref{7.10}) and
(\ref{7.9a}) reconstruct coefficients $c_{\alpha}(t_{1},t_{2},t_{3})$ of the
case (\ref{3.1}), and the map (\ref{mapka}) becomes:
\begin{align}
a_{7} &  =\frac{1}{2}b_{4}^{2},\quad a_{6}=b_{3}b_{4},\quad a_{5}=\frac{1}%
{2}b_{3}^{2}+b_{2}b_{4},\quad a_{4}=b_{1}b_{4}+b_{2}b_{3},\quad
\label{gwiazdka}\\
a_{3} &  =\frac{1}{2}b_{2}^{2}+b_{0}b_{4}+b_{1}b_{3}+\overline{b}-b_{4},\quad
a_{-1}=\frac{1}{2}b_{0}^{2}.\nonumber
\end{align}

\end{example}

\begin{example}
Finally, for $n=3$ and $m=4$, as again $H_{r}^{B}=h_{r}^{B}$ for all $r$ so,
according to (\ref{7.9b}) and (\ref{7.9c}) we get the following set of first
order PDE's%
\[%
\begin{array}
[c]{lll}%
\dfrac{\partial d_{3}}{\partial t_{k}}=0, & \ \dfrac{\partial d_{0}}{\partial
t_{k}}=0, & \ k=1,2,3,\\
& \  & \\
\dfrac{\partial d_{4}}{\partial t_{1}}=d_{4}, & \ \dfrac{\partial d_{4}%
}{\partial t_{2}}=0, & \ \dfrac{\partial d_{4}}{\partial t_{3}}=0,\\
& \  & \\
\dfrac{\partial d_{1}}{\partial t_{1}}=0, & \ \dfrac{\partial d_{1}}{\partial
t_{2}}=2d_{0}, & \ \dfrac{\partial d_{1}}{\partial t_{3}}=0,\\
& \  & \\
\dfrac{\partial d_{2}}{\partial t_{1}}=0, & \ \dfrac{\partial d_{2}}{\partial
t_{2}}=d_{1}, & \ \dfrac{\partial d_{2}}{\partial t_{3}}=d_{0},
\end{array}
\]
which are solved to
\[
d_{0}(t)=b_{0},\quad d_{3}(t)=b_{3},\quad d_{4}(t)=b_{4}\exp(t_{1}),\quad
d_{1}(t)=2b_{0}t_{2}+b_{1},\quad d_{2}(t)=b_{0}(t_{3}+t_{2}^{2})+b_{1}%
t_{2}+b_{2},
\]
obtained previously in (\ref{v4}). The formulas (\ref{7.10}) and (\ref{7.9a})
reconstruct the coefficients $c_{\alpha}(t_{1},t_{2},t_{3})$ of the case
(\ref{3.4}), yielding the map (\ref{mapka}) given by
\[
a_{4}=\frac{1}{2}b_{4}^{2},\quad a_{3}=-\frac{1}{2}b_{4}^{2}+b_{3}%
b_{4}+\overline{b}-b_{4},\quad a_{-1}=b_{0}b_{3}+b_{1}b_{2},\quad a_{-2}%
=\frac{1}{2}b_{1}^{2}+b_{2}b_{0},\quad a_{-3}=b_{1}b_{0},\quad a_{-4}=\frac
{1}{2}b_{0}^{2}.
\]

\end{example}

An elementary calculation using Lemma \ref{VDM} shows that in the Vi\`{e}te
coordinates (\ref{26}) the transformation (\ref{7.3}) takes the form
\begin{equation}
q_{i}=q_{i}^{\prime},\qquad p_{i}=p_{i}^{\prime}+\sum_{\gamma=0}%
^{n+1}d_{\gamma}(t_{1},\dotsc,t_{n})V_{1}^{(n-m-i+\gamma)}. \label{7.11}%
\end{equation}
Also the functions $Z_{k}$ are expressible by basic separable potentials (see
\ref{103}).

\begin{example}
\label{4e}Let us relate the Hamiltonian system from Example \ref{3e} with the
Hamiltonian system from Example \ref{1e}. For $n=3$, $m=1$ and $b_{0}%
=b_{1}=b_{2}=b_{4}=\overline{b}=0$, $b_{3}\in\mathbb{R}$, the relation
(\ref{w1}) becomes
\[
d_{0}(t)=0,\quad d_{1}=b_{3}(t_{2}+t_{3}^{2}),\quad d_{2}(t)=2b_{3}t_{3},\quad
d_{3}(t)=b_{3},\quad d_{4}(t)=0.
\]
Moreover, in accordance with (\ref{103})
\[
Z_{1}=-q_{1},\quad Z_{2}=q_{1}^{2}-2q_{2},\quad Z_{3}=-q_{1}^{3}+3q_{1}%
q_{2}-3q_{3},
\]
so that (\ref{7.9}) and (\ref{7.12}) become
\begin{gather*}
\frac{\partial F}{\partial t_{1}}=\mathcal{Z}_{1}=d_{4}Z_{1}=0,\quad
\frac{\partial F}{\partial t_{2}}=\mathcal{Z}_{2}=d_{3}Z_{1}+d_{4}Z_{2}%
=-b_{3}q_{1},\\
\frac{\partial F}{\partial t_{3}}=\mathcal{Z}_{3}+t_{2}\mathcal{Z}_{1}%
=d_{3}Z_{2}+d_{2}Z_{1}=b_{3}(q_{1}^{2}-2q_{2})-2b_{3}t_{3}q_{1},
\end{gather*}
while the transformation (\ref{7.11}) specifies to
\begin{align*}
q_{i} &  =q_{i}^{\prime},\quad i=1,2,3,\\
p_{1} &  =p_{1}^{\prime}+b_{3}V_{1}^{(4)}+2b_{3}t_{3}V_{1}^{(3)}+b_{3}%
(t_{2}+t_{3}^{2})V_{1}^{(2)}=p_{1}^{\prime}+b_{3}(q_{2}-q_{1}^{2})+2b_{3}%
t_{3}q_{1}-b_{3}(t_{2}+t_{3}^{2}),\\
p_{2} &  =p_{2}^{\prime}+b_{3}V_{1}^{(3)}+2b_{3}t_{3}V_{1}^{(2)}+b_{3}%
(t_{2}+t_{3}^{2})V_{1}^{(1)}=p_{2}^{\prime}+b_{3}q_{1}-2b_{3}t_{3},\\
p_{3} &  =p_{3}^{\prime}-b_{3}.
\end{align*}
Then, according to our theory (cf. (\ref{7.5})) and provided that (see
(\ref{gwiazdka})) $a_{5}=\frac{1}{2}b_{3}^{2}$:
\[
H_{r}^{A}(q,p^{\prime},t)=H_{r}^{B}(q,p^{\prime},t)+\frac{\partial
F(q,p^{\prime},t)}{\partial t_{r}},\quad r=1,\ldots,n
\]
while the non-dynamical parameters in $H_{r}^{A}$ are given by
\[
c_{0}=2b_{3}(t_{2}+t_{3}^{2})(b_3t_{2}t_{3}+1),\quad
c_{1}=\frac{1}{2}b_{3}^{2}(t_{2}^{2}+t_{3}^{4}+2t_{2}t_{3}^{2})+2b_{3}%
t_{3},\quad c_{2}=2b_{3}^{2}(t_{2}+t_{3}^{2})t_{3},
\]
which is another particular solution of (\ref{sol}) in Example \ref{1e}.
\end{example}

As we mentioned above, the map $(b_{0},\dotsc,b_{n+1},\overline{b}%
)\rightarrow(a_{-m},\dotsc,a_{-1},a_{n},\dotsc,a_{2n-m+2})$ is not bijective,
not every system with ordinary potential has a representation with magnetic
potential. To illustrate such a case, let us consider the non-autonomous
H\'{e}non-Heiles system from Example \ref{2e}. For the case $n=2$ and $m=1$ we
get
\[
d_{3}(t)=b_{3},\quad d_{2}(t)=b_{2}+2b_{3}t_{2},\quad d_{1}(t)=b_{1}%
+b_{2}t_{2}+b_{3}(t_{1}+t_{2}^{2}),\quad d_{0}=b_{0}%
\]
and the map (\ref{mapka}) is
\begin{equation}
a_{5}=\frac{1}{2}b_{3}^{2},\quad a_{4}=b_{2}b_{3},\quad a_{3}=\frac{1}{2}%
b_{2}^{2}+b_{1}b_{3},\quad a_{2}=b_{0}b_{3}+b_{1}b_{2}+\overline{b}%
-b_{3},\quad a_{-1}=\frac{1}{2}b_{0}. \label{HH}%
\end{equation}
For the non-autonomous H\'{e}non-Heiles system in Example \ref{2e} we have
$a_{5}=a_{3}=a_{2}=0$, $a_{4}=-1$, $a_{-1}=-\frac{1}{4}\alpha$ and for such a
choice the system (\ref{HH}) has no solutions for $(b_{0},\dotsc
,b_{3},\overline{b})$. In consequence, the non-autonomous H\'{e}non-Heiles
system has no equivalent representation with magnetic potentials.

\begin{remark}
\label{r2}Let us observe that for non-autonomous Hamiltonians $H_{k}$, $k\in
I_{2}^{m}$, the evolution parameter $t_{n+2-m}$ enters through the exponential
function. Thus we can introduce new evolution parameter
\[
t_{n+2-m}^{\prime}=\exp(t_{n+2-m})\rightarrow\frac{d}{dt_{n+2-m}}%
=t_{n+2-m}^{\prime}\frac{d}{dt_{n+2-m}^{\prime}}%
\]
so that
\[
\xi_{t_{n+2-m}^{\prime}}=Y_{n+2-m}(\xi,t_{1},\dotsc,t_{n+2-m}^{\prime}%
,\dotsc,t_{n})=\pi d\left(  \frac{1}{t_{n+2-m}^{\prime}}H_{r}(\xi,t_{1}%
,\dotsc,t_{n+2-m}^{\prime},\dotsc,t_{n})\right)  .
\]

\end{remark}

\section{One-dimensional magnetic systems and Painlev\'{e} equations}

We will now demonstrate how the trivial case $n=1$ with $m=0,1,2$ leads to the
well known Painlev\'{e} equations $P_{I}-P_{IV}$. The magnetic separation
relations (\ref{7.1}) take in this case the simple form
\[
e(t)\lambda_{1}+\left[  d_{0}(t)+d_{1}(t)\lambda_{1}+d_{2}(t)\lambda_{1}%
^{2}\right]  \mu_{1}+h^{B}=\frac{1}{2}\lambda_{1}^{m}\mu_{1}^{2}%
\]
with $m$ being either $1,2$ or $3$ (note that the quasi-St\"{a}ckel term is
absent now) and thus, the Hamiltonian with magnetic potential written in the
Vi\`{e}te coordinates $\lambda_{1}=-q$ and $\mu_{1}=p$ is
\begin{equation}
h^{B}=\frac{1}{2}(-q)^{m}p^{2}-\left[  d_{0}(t)-d_{1}(t)q+d_{2}(t)q^{2}%
\right]  p+e(t)q. \label{8.1}%
\end{equation}
The formulas (\ref{7.9})\ and (\ref{7.12}) reduce to
\[
\frac{dF}{dt}=d_{2}Z_{1}=-d_{2}q
\]
while the set of PDE's (\ref{7.9b})--(\ref{7.9c}) in Theorem \ref{PDE} has the
solution
\begin{align*}
&  \text{for $m=0$:} & d_{2}(t)  &  =b_{2}, & d_{1}(t)  &  =b_{1}, & d_{0}(t)
&  =b_{2}t+b_{0}, & \text{with }e(t)  &  =\overline{b},\\
&  \text{for $m=1$:} & d_{2}(t)  &  =b_{2}, & d_{0}(t)  &  =b_{0}, & d_{1}(t)
&  =b_{2}t+b_{1}, & \text{with }e(t)  &  =\overline{b},\\
&  \text{for $m=2$:} & d_{0}(t)  &  =b_{0}, & d_{1}(t)  &  =b_{1}, & d_{2}(t)
&  =b_{2}\exp(t), & \text{with }e(t)  &  =\overline{b}\exp(t).
\end{align*}

Let us now compute, with the help of Theorem \ref{PDE} and the transformation
formula (\ref{7.5}), the non-magnetic representations of the Hamiltonian
(\ref{8.1}). Consider first the case $m=0$. Applying in this case the
transformation formula (\ref{7.5}) to (\ref{8.1}) we get (up to terms
independent on $q$ and $p$), the non-magnetic Hamiltonian
\[
h^{A}=\frac{1}{2}p^{2}-a_{4}q^{4}+a_{3}q^{3}-(2a_{4}t+a_{2})q^{2}%
+(a_{3}t+a_{1})q
\]
where, according with our theory%
\[
a_{4}=\frac{1}{2}b_{2}^{2},\quad a_{3}=b_{1}b_{2},\quad a_{2}=\frac{1}{2}%
b_{1}^{2}+b_{0}b_{2},\quad a_{1}=b_{0}b_{1}+\overline{b}-b_{2}.
\]
Eliminating $p$ from the corresponding Hamiltonian equations of motion we get
\begin{equation}
q_{tt}=4a_{4}q^{3}-3a_{3}q^{2}+2(2a_{4}t+a_{2})q-(a_{3}t+a_{1}). \label{kasia}%
\end{equation}
For $a_{4}=a_{2}=a_{1}=0$ and $a_{3}=-1$ this equation reduces to
\begin{equation}
q_{tt}=3q^{2}+t, \label{P1}%
\end{equation}
which after rescaling $q\rightarrow2^{\frac{3}{5}}q$, $t\rightarrow2^{\frac
{1}{5}}t$ becomes the Painlev\'{e} I equation%
\[
q_{tt}=6q^{2}+t,
\]
while for $a_{4}=\frac{1}{4}$, $a_{3}=a_{2}=0$ and $a_{1}=-\alpha$ the
equation (\ref{kasia}) reduces to
\begin{equation}
q_{tt}=q^{3}+tq+\alpha, \label{P2}%
\end{equation}
which up to the rescaling $q\rightarrow2^{\frac{1}{2}}q$, $\alpha
\rightarrow2^{-\frac{1}{2}}\alpha$ is the Painlev\'{e} II equation%
\[
q_{tt}=2q^{3}+tq+\alpha.
\]
Observe that from (\ref{7.10}) it follows that Painlev\'{e} I equation
(\ref{P1}) does not have any representation with magnetic potential, while
Painlev\'{e} II equation (\ref{P2}) has the magnetic representation
(\ref{8.1}) with $b_{0}=b_{1}=0$, $b_{2}=2^{\frac{1}{2}}$, $\overline
{b}=\alpha$.

For $m=1$, the corresponding non-magnetic Hamiltonian attains the form
\begin{equation}
h^{A}=-\frac{1}{2}qp^{2}+a_{3}q^{3}-(2a_{3}t+a_{2})q^{2}+(a_{3}t^{2}%
+a_{2}t+a_{1})q+a_{-1}q^{-1}, \label{8.6}%
\end{equation}
(again up to terms independent on $q$ and $p$) where%
\begin{equation}
a_{3}=\frac{1}{2}b_{2}^{2},\quad a_{2}=b_{1}b_{2},\quad a_{1}=\frac{1}{2}%
b_{1}^{2}+b_{0}b_{2}+\overline{b}-b_{2},\quad a_{-1}=\frac{1}{2}b_{0}^{2}.
\label{8.7}%
\end{equation}
Eliminating $p$ from Hamiltonian equations of motion we obtain
\begin{equation}
qq_{tt}=\frac{1}{2}q_{t}^{2}+3a_{3}q^{4}-2(2a_{3}t+a_{2})q^{3}+(a_{3}%
t^{2}+a_{2}t+a_{1})q^{2}-a_{-1}. \label{8.8}%
\end{equation}
For $a_{3}=1$ and $a_{2}=0$ the equation (\ref{8.8}) reduces to
\begin{equation}
qq_{tt}=\frac{1}{2}q_{t}^{2}+3q^{4}-4tq^{3}+(t^{2}+a_{1})q^{2}-a_{-1}
\label{8.9}%
\end{equation}
which after rescaling $t\rightarrow2^{\frac{1}{4}}t$, $q\rightarrow
-2^{-\frac{3}{4}}q$, $2^{-\frac{1}{2}}a_{1}=\alpha$ and $4a_{-1}=\beta$ is the
Painlev\'{e} IV equation%
\[
qq_{tt}=\frac{1}{2}q_{t}^{2}+\frac{3}{2}q^{4}+4tq^{3}+2(t^{2}-\alpha
)q^{2}+\beta.
\]
Observe that from (\ref{8.7}) it follows that the equation (\ref{8.9}), and
thus so Painlev\'{e} IV, has the magnetic representation (\ref{8.1}) for
$b_{1}=0$ and $\frac{1}{2}b_{2}^{2}=1$.

Finally, for $m=2$, the corresponding non-magnetic Hamiltonian becomes
\[
h^{A}=\frac{1}{t^{\prime}}\left(  \frac{1}{2}q^{2}p^{2}-a_{2}t^{\prime2}%
q^{2}+a_{1}t^{\prime}q+a_{-1}q^{-1}-a_{-2}q^{-2}\right)  ,
\]
where
\[
a_{2}=\frac{1}{2}b_{2}^{2},\quad a_{1}=b_{1}b_{2}+\overline{b}-b_{2},\quad
a_{-1}=b_{0}b_{1},\quad a_{-1}=\frac{1}{2}b_{0}^{2}%
\]
and
\begin{equation}
t^{\prime}qq_{t^{\prime}t^{\prime}}=t^{\prime}q_{t^{\prime}}^{2}%
-qq_{t^{\prime}}+2a_{2}t^{\prime}q^{4}-a_{1}q^{3}+a_{-1}\frac{1}{t^{\prime}%
}q-2a_{-2}\frac{1}{t^{\prime}}, \label{12}%
\end{equation}
where, according to Remark \ref{r2}, $t^{\prime}=\exp(t)$. The transformation
\[
t^{\prime}=\frac{1}{2}\tau^{2},\qquad q=\frac{1}{\tau}w,
\]
turns (\ref{12}) to the form
\[
\tau ww_{\tau\tau}=\tau w_{\tau}^{2}-ww_{\tau}+2a_{2}\tau w^{4}-2a_{1}%
w^{3}+4a_{-1}w-8a_{-2}\tau
\]
which is exactly the Painlev\'{e} III equation
\[
\tau ww_{\tau\tau}=\tau w_{\tau}^{2}-ww_{\tau}+\gamma\tau w^{4}+\alpha
w^{3}+\beta w+\delta\tau
\]
with
\[
-8a_{-2}=\delta,\quad4a_{-1}=\beta,\quad-2a_{1}=\alpha,\quad2a_{2}=\gamma.
\]
Thus, within our formalism, we have rediscovered the first four one-field
Painlev\'{e} equations $P_{I}-P_{IV}$.

\appendix

\setcounter{equation}{0} \renewcommand{\theequation}{A.\arabic{equation}}

\section*{Appendix A}

In this Appendix we prove Theorem \ref{1t}. We prove only the case (i) of the
theorem as other cases can be treated similarly. We start by proving
(\ref{t2}), that is, we will prove that for any $r,s\in\{1\}\cup I_{1}%
^{m}=\{1,2,\ldots,n-m+1\}$ and for any $k=1,\dotsc,m$
\begin{equation}
\left\{  h_{r}^{(-k)},h_{s}^{(-k)}\right\}  =(s-r)h_{r+s+m-n-2}^{(-k)},
\label{A1}%
\end{equation}
where
\[
h_{r}^{(-k)}=h_{r}+V_{r}^{(-k)}=E_{r}+W_{r}+V_{r}^{(-k)}.
\]
Note first that $0<m\leq n-1$ in this case as for $m=0$ there is no $k$. We
have, due to (\ref{str}) and since the St\"{a}ckel Hamiltonians $E_{r}%
+V_{r}^{(-k)}$ and $E_{s}+V_{s}^{(-k)}$ Poisson commute,
\begin{align*}
\left\{  h_{r}^{(-k)},h_{s}^{(-k)}\right\}   &  = \left\{  h_{r}+V_{r}%
^{(-k)},h_{s}+V_{s}^{(-k)}\right\}  =\left\{  E_{r}+W_{r}+V_{r}^{(-k)}%
,E_{s}+W_{s}+V_{s}^{(-k)}\right\} \\
&  = (s-r)\left(  E_{r+s+m-n-2}+W_{r+s+m-n-2}\right)  +\left\{  W_{r}%
,V_{s}^{(-k)}\right\}  +\left\{  V_{r}^{(-k)},W_{s}\right\}
\end{align*}
so in order to prove (\ref{A1}) it suffices to prove that for any
$r,s\in\{1\}\cup I_{1}^{m}=\{1,2,\ldots,n-m+1\}$ and for any $k=1,\dotsc,m$%
\begin{equation}
\left\{  W_{r},V_{s}^{(-k)}\right\}  +\left\{  V_{r}^{(-k)},W_{s}\right\}
=(s-r)V_{r+s+m-n-2}^{(-k)}. \label{A2}%
\end{equation}
Note first that the following relations hold:

for $k=1,2,\ldots$
\begin{equation}
\frac{\partial V_{r}^{(-k)}}{\partial q_{i}}=\frac{\partial V_{r}^{(-k+1)}%
}{\partial q_{i+1}}+\delta_{r,i+1}V_{1}^{(-k)},\quad i=1,2,\dotsc,n-1,
\label{A3}%
\end{equation}

for $r=1,\dotsc,n$ and $k\in\mathbb{Z}$%
\begin{equation}
V_{r}^{(-k)}=V_{r-1}^{(-k+1)}-\frac{q_{r-1}}{q_{n}}V_{n}^{(-k+1)} \label{A4}%
\end{equation}

and%
\begin{equation}
V_{1}^{(-k)}=-\frac{1}{q_{n}}V_{n}^{(-k+1)}, \label{A5}%
\end{equation}
where of course (\ref{A5}) follows from (\ref{A4}). The recursion (\ref{A4})
can be reversed yielding for $r=1,\dotsc,n$ and $k\in\mathbb{Z}$
\begin{equation}
V_{r}^{(k+1)}=V_{r+1}^{(k)}-q_{r}V_{1}^{(k)}. \label{A5.5}%
\end{equation}

\bigskip

Further%
\begin{equation}
\frac{\partial W_{r}}{\partial p_{i}}=(n+1-m-i)q_{r+i+m-n-2},\quad\text{where
$q_{j}=0$ if $j>n$ or $j<0$}. \label{A6}%
\end{equation}
For $r,s\in I_{1}^{m}=\{2,\ldots,n-m+1\}$ and $0<m<n-1$ we will prove
(\ref{A2}) by induction. Since in the process of induction we will pass
between systems with different $m$, we will in the rest of the proof denote
$W_{r}$ with a given $m$ by $W_{m,r}$ and likewise $h_{r}^{(-k)}$ with a given
$m$ by $h_{m,r}^{(-k)}$ and $h_{r}^{(n+k)}$ with a given $m$ by $h_{m,r}%
^{(n+k)}$. We start from the easy proved relations for $k=1$ and $0<m<n-1$
\[
\left\{  V_{r}^{(-1)},W_{m,s}\right\}  +\left\{  W_{m,r},V_{s}^{(-1)}\right\}
=(s-r)V_{r+s+m-n-2}^{(-1)}.
\]
Assuming now that (\ref{A2}) is true for a pair $(k,m)$ such that $k\leq m\leq
n-2$, we will prove that the same formula is valid for the pair $(k+1,m+1)$
for all \thinspace$r,s\in I_{1}^{m+1}$, which will prove that (\ref{A2}) is
valid for any $r,s\in I_{1}^{m}=\{2,\ldots,n-m+1\}$ and for any $k=1,\dotsc
,m$. The induction terminates when $m=n-2$. Observe also that the functions
$W_{m,s}$ do not depend on $p_{n-m+1},\dotsc,p_{n}$. Then, for $r,s=2,\dotsc
,n-m$
\begin{align*}
&  \left\{  V_{r}^{(-k-1)},W_{m+1,s}\right\}  +\left\{  W_{m+1,r}%
,V_{s}^{(-k-1)}\right\} \\
&  \overset{(\ref{A6})}{=}\sum_{i=1}^{n-(m+1)}(n-m-i)\left[  q_{s+i+m-n-1}%
\frac{\partial V_{r}^{(-k-1)}}{\partial q_{i}}-q_{r+i+m-n-1}\frac{\partial
V_{s}^{(-k-1)}}{\partial q_{i}}\right] \\
&  \overset{(\ref{A3})}{=}\sum_{i=1}^{n-(m+1)}(n-m-i)\left[  q_{s+i+m-n-1}%
\left(  \frac{\partial V_{r}^{(-k)}}{\partial q_{i+1}}+\delta_{r,i+1}%
V_{1}^{(-k-1)}\right)  \right. \\
&  \omit\hfill$\displaystyle \left.  {}-q_{r+i+m-n-1}\left(
\frac{\partial V_s^{(-k)}}{\partial
q_{i+1}}+\delta_{s,i+1}V_1^{(-k-1)}\right)  \right]
$\ignorespaces\\
&  \overset{\phantom{(\ref{A6})}}{=} \sum_{i=1}^{n-(m+1)}(n-m-i)\left[
q_{s+i+m-n-1}\frac{\partial V_{r}^{(-k)}}{\partial q_{i+1}}-q_{r+i+m-n-1}%
\frac{\partial V_{s}^{(-k)}}{\partial q_{i+1}}\right]  +(s-r)q_{r+s+m-n-2}%
V_{1}^{(-k-1)}\\
&  \overset{\phantom{(\ref{A6})}}{=} \sum_{i=2}^{n-m}(n+1-m-i)\left[
q_{s+i+m-n-2}\frac{\partial V_{r}^{(-k)}}{\partial q_{i}}-q_{r+i+m-n-2}%
\frac{\partial V_{s}^{(-k)}}{\partial q_{i}}\right]  +(s-r)q_{r+s+m-n-2}%
V_{1}^{(-k-1)}\\
&  \overset{\phantom{(\ref{A6})}}{=} \left\{  V_{r}^{(-k)},W_{m,s}\right\}
+\left\{  W_{m,r},V_{s}^{(-k)}\right\} \\
&  \omit\hfill$\displaystyle {} -(n-m)\left[  \overset{=0}{q_{s+m-n-1}}%
\frac{\partial V_r^{(-k)}}{\partial
q_1}-\overset{=0}{q_{r+m-n-1}}\frac{\partial
V_s^{(-k)}}{\partial q_1}\right]  +(s-r)q_{r+s+m-n-2}V_1^{(-k-1)}$\ignorespaces\\
&  \overset{(\ref{A2})}{=}(s-r)\left[  V_{r+s+m-n-2}^{(-k)}+q_{r+s+m-n-2}%
V_{1}^{(-k-1)}\right]  \overset{(\ref{A5})}{=}(s-r)\left[  V_{r+s+m-n-2}%
^{(-k)}-\frac{q_{r+s+m-n-2}}{q_{n}}V_{n}^{(-k)}\right] \\
&  \overset{(\ref{A4})}{=}(s-r)V_{r+s+m-n-1}^{(-k-1)},
\end{align*}
thus (\ref{A2}) is valid for $(k+1,m+1)$ which concludes the inductive proof.
The fact that $q_{s+m-n-1}=0$ in the third line from below is since
$s+m-n-1\leq-1$. For the same reasons we have $q_{r+m-n-1}=0$. Finally we
have
\[
\left\{  h_{m,1}^{(-k)},h_{m,s}^{(-k)}\right\}  =\left\{  V_{1}^{(-k)}%
,W_{m,s}\right\}  =0,\quad1\leq k\leq m,\quad s\in I_{1}^{m}%
\]
as $V_{1}^{(-k)}=V_{1}^{(-k)}(q_{n},\dotsc,q_{n-k+1})$ and $W_{m,s}$ functions
do not depend on $p_{n-m+1},\dotsc,p_{n}$.

Let us now prove (\ref{t1}), i.e. for any $r,s\in I_{1}^{m}$ and for any
$k=0,\dotsc,n-m+2$ we have
\begin{align}
\left\{  h_{m,r}^{(n+k)},h_{m,s}^{(n+k)}\right\}   &  = (s-r)h_{m,r+s+m-n-2}%
^{(n+k)}\nonumber\\
&  \quad{} +(2r+k+m-n-2)V_{s}^{(r+k+m-2)}-(2s+k+m-n-2)V_{r}^{(s+k+m-2)}.
\label{A8}%
\end{align}
By a reasoning analogous to the one above, the proof of (\ref{A8}) boils down
to proving that the following relation is valid for any $r,s\in\{1\}\cup
I_{1}^{m}$ and for any $k=0,\dotsc,n-m+2$:
\begin{align}
\left\{  V_{r}^{(n+k)},W_{m,s}\right\}  +\left\{  W_{m,r},V_{s}^{(n+k)}%
\right\}   &  = (s-r)V_{r+s+m-n-2}^{(n+k)}+(2r+m+k-n-2)V_{s}^{(r+m+k-2)}%
\nonumber\\
&  \quad{} -(2s+m+k-n-2)V_{r}^{(s+m+k-2)}. \label{A9}%
\end{align}
Again, we will proceed by induction. We start by noting that the formula
(\ref{A9}) is easily proved for $k=0,\dotsc,4$, $0\leq m\leq n-1$ and
$r,s\in\{1\}\cup I_{1}^{m}=\{1,2,\dotsc,n-m+1\}$. Assuming now that (\ref{A9})
is true for a pair $(k,m)$ such that $0\leq k\leq n-m+2$ we prove below that
the same formula is valid for the pair $(k-1,m+1)$ and for $r,s\in\{1\}\cup
I_{1}^{m+1}$. The induction terminates if either $k=0$ or $m=n-1$. We also
remind that $W_{m,s}$ do not depend on $p_{n-m+1},\dotsc,p_{n}$. So, for
$r,s\in\{1\}\cup I_{1}^{m+1}$, $m\leq n-2$ and $k\leq n-m+2$
\begin{align*}
&  \left\{  V_{r}^{(n+k-1)},W_{m+1,s}\right\}  +\left\{  W_{m+1,r}%
,V_{s}^{(n+k-1)}\right\} \\
&  \overset{(\ref{A6})}{=}\sum_{i=1}^{n-(m+1)}(n-m-i)\left[  q_{s+i+m-n-1}%
\frac{\partial V_{r}^{(n+k-1)}}{\partial q_{i}}-q_{r+i+m-n-1}\frac{\partial
V_{s}^{(n+k-1)}}{\partial q_{i}}\right] \\
&  \overset{(\ref{A3})}{=}\sum_{i=1}^{n-(m+1)}(n-m-i)\left[  q_{s+i+m-n-1}%
\left(  \frac{\partial V_{r}^{(n+k)}}{\partial q_{i+1}}+\delta_{i+1,r}%
V_{1}^{(n+k-1)}\right)  \right. \\
&  \omit\hfill$\displaystyle {} -\left.  q_{r+i+m-n-1}\left(
\frac{\partial V_s^{(n+k)}}{\partial
q_{i+1}}+\delta_{i+1,s}V_1^{(n+k-1)}\right)  \right]
$\ignorespaces\\
&  \overset{\phantom{(\ref{A6})}}{=} \sum_{i=1}^{n-(m+1)}(n-m-i)\left[
q_{s+i+m-n-1}\frac{\partial V_{r}^{(n+k)}}{\partial q_{i+1}}-q_{r+i+m-n-1}%
\frac{\partial V_{s}^{(n+k)}}{\partial q_{i+1}}\right]  +(s-r)q_{r+s+m-n-2}%
V_{1}^{(n+k-1)}\\
&  \overset{\phantom{(\ref{A6})}}{=} \sum_{i=2}^{n-m}(n+1-m-i)\left[
q_{s+i+m-n-2}\frac{\partial V_{r}^{(n+k)}}{\partial q_{i}}-q_{r+i+m-n-2}%
\frac{\partial V_{s}^{(n+k)}}{\partial q_{i}}\right]  +(s-r)q_{r+s+m-n-2}
V_{1}^{(n+k-1)}\\
&  \overset{\phantom{(\ref{A6})}}{=} \left\{  V_{r}^{(n+k)},W_{m,s}\right\}
+\left\{  W_{m,r},V_{s}^{(n,n+k)}\right\} \\
&  \omit\hfill$\displaystyle {} -(n-m)\left[  \overset{=0}{q_{s+m-n-1}}%
\frac{\partial V_r^{(n+k)}}{\partial
q_1}-\overset{=0}{q_{r+m-n-1}}\frac{\partial
V_s^{(n+k)}}{\partial q_1}\right]  +(s-r)q_{r+s+m-n-2} V_1^{(n+k-1)}$\ignorespaces\\
&  \overset{(\ref{A9})}{=}(s-r)V_{r+s+m-n-2}^{(n+k)}+(2r+m+k-n-2)V_{s}%
^{(r+m+k-2)}-(2s+m+k-n-2)V_{r}^{(s+m+k-2)}\\
&  \omit\hfill$\displaystyle {} +(s-r)q_{r+s+m-n-2} V_1^{(n+k-1)}$\ignorespaces\\
&  \overset{\phantom{(\ref{A6})}}{=} (s-r)\left[  V_{r+s+m-n-2}^{(n+k)}%
+q_{r+s+m-n-2} V_{1}^{(n+k-1)}\right]  +(2r+m+k-n-2)V_{s}^{(r+m+k-2)}\\
&  \omit\hfill$\displaystyle {} -(2s+m+k-n-2)V_r^{(s+m+k-2)}$\ignorespaces\\
&  \overset{(\ref{A5.5})}{=}(s-r)V_{r+s+m-n-1}^{(n+k-1)}+(2r+m+k-n-2)V_{s}%
^{(r+m+k-2)}-(2s+m+k-n-2)V_{r}^{(s+m+k-2)}%
\end{align*}
which is exactly (\ref{A9}) for the pair $(k-1,m+1)$. Again, $q_{s+m-n-1}=0$
since $s+m-n-1\leq-1$. This concludes the proof.

\setcounter{equation}{0} \renewcommand{\theequation}{B.\arabic{equation}}

\section*{Appendix B}

The proof of (\ref{100a}) is immediate. In what follows we will present the
proof of (\ref{100b}). Formula (\ref{100c}) can be proved analogically. With
the help of (8.5) we have that
\[
S_{r}=W_{r}-W_{r}^{\prime}=\sum_{i=1}^{n}(\mu_{i}-\mu_{i}^{\prime})J_{r}%
^{i}=\sum_{i=1}^{n}\sum_{\gamma=0}^{n+1}J_{r}^{i}d_{\gamma}\lambda_{i}%
^{\gamma-m},
\]
where according to (\ref{j1})
\[
J_{r}^{i}=-\sum_{k=1}^{r-1}k\rho_{r-k-1}\frac{\lambda_{i}^{m+k-1}}{\Delta_{i}%
},\quad r\in I_{1}^{m}.
\]
Thus
\begin{equation}
S_{r}=-\sum_{i=1}^{n}\sum_{\gamma=0}^{n+1}\sum_{k=1}^{r-1}k\rho_{r-k-1}%
d_{\gamma}\frac{\lambda_{i}^{\gamma+k-1}}{\Delta_{i}}. \label{101}%
\end{equation}

Next, observe that by virtue of (\ref{Vnew}) the basic potentials
$V_{k}^{(\gamma)}$ can be written in the following way
\[
V_{k}^{(\gamma)}=\sum_{i=1}^{n}\frac{\partial\rho_{k}}{\partial\lambda_{i}%
}\frac{\lambda_{i}^{\gamma}}{\Delta_{i}}=-\sum_{i=1}^{n}\sum_{s=0}^{k-1}%
\rho_{s}\frac{\lambda_{i}^{\gamma+k-s-1}}{\Delta_{i}},
\]
where we have used the first identity in (\ref{potrzebne}):
\[
\frac{\partial\rho_{k}}{\partial\lambda_{i}}=-\sum_{s=0}^{k-1}\rho_{s}%
\lambda_{i}^{k-s-1}.
\]
Thus, the formula (\ref{101}) takes the form
\begin{align}
S_{r}  &  =-\sum_{i=1}^{n}\sum_{\gamma=0}^{n+1}\sum_{s=0}^{r-2}%
(r-s-1)d_{\gamma}\rho_{s}\frac{\lambda_{i}^{\gamma+r-s-2}}{\Delta_{i}}%
=-\sum_{i=1}^{n}\sum_{\gamma=0}^{n+1}\sum_{k=1}^{r-1}\sum_{s=0}^{k-1}%
d_{\gamma}\rho_{s}\frac{\lambda_{i}^{\gamma+r-s-2}}{\Delta_{i}}=\sum
_{\gamma=0}^{n+1}\sum_{k=1}^{r-1}d_{\gamma}V_{k}^{(\gamma+r-k-1)}\nonumber\\
&  =\sum_{\gamma=-n+r-1}^{r}d_{\gamma+n-r+1}\sum_{k=1}^{r-1}V_{k}%
^{(n+\gamma-k)}, \label{102}%
\end{align}
where we have used the identity
\[
\sum_{s=0}^{r-2}(r-s-1)a_{s}=\sum_{k=1}^{r-1}\sum_{s=0}^{k-1}a_{s}.
\]

From (\ref{7a}) we get that $V_{k}^{(n-k)}=-1$ for $k=1,\dotsc,n$ and
$V_{k}^{(n+\gamma-k)}=0$ for $k=1,\dotsc,r-1$ and $\gamma=-n+r-1,\dotsc,-1 $
or $k=1,\dotsc,n$ and $\gamma=1,\dotsc,k-1$. With the help of these formulas
the equation (\ref{102}) can be written in the form
\begin{align*}
S_{r}  &  =-(r-1)d_{n-r+1}+\sum_{\gamma=1}^{r}d_{\gamma+n-r+1}\sum_{k=1}%
^{r-1}V_{k}^{(n+\gamma-k)}\\
&  =-(r-1)d_{n-r+1}+\sum_{\gamma=1}^{r}d_{\gamma+n-r+1}\sum_{k=1}^{r}%
V_{k}^{(n+\gamma-k)}-\sum_{\gamma=1}^{r}d_{\gamma+n-r+1}V_{r}^{(n+\gamma-r)}\\
&  =-(r-1)d_{n-r+1}+\sum_{\gamma=1}^{r}d_{\gamma+n-r+1}\sum_{k=1}^{\gamma
}V_{k}^{(n+\gamma-k)}-d_{n+1}V_{r}^{(n)}.
\end{align*}
Assuming that
\begin{equation}
\sum_{k=1}^{\gamma}V_{k}^{(n+\gamma-k)}=-Z_{\gamma},\quad\gamma=1,\dotsc,r
\label{103}%
\end{equation}
we get (\ref{100b}). In what follows we will show that formula (\ref{103})
indeed holds.

We will prove (\ref{103}) by induction. For $\gamma=1$ this formula clearly
holds since
\[
V_{1}^{(n)}=\rho_{1}=-\lambda_{1}-\dotsb-\lambda_{n}=-Z_{1}.
\]
Now, let us fix $s\in\{2,3,\dotsc\}$ and assume that formula (\ref{103}) holds
for every $\gamma<s$. We will show that it holds for $\gamma=s$. Using the
recursive formula (\ref{A5.5})
\begin{equation}
V_{k}^{(\alpha)}=V_{k+1}^{(\alpha-1)}-\rho_{k}V_{1}^{(\alpha-1)} \label{104}%
\end{equation}
we get that
\begin{align*}
\sum_{k=1}^{s}V_{k}^{(n+s-k)}  &  =V_{1}^{(n+s-1)}+V_{2}^{(n+s-2)}%
+V_{3}^{(n+s-3)}+\dotsb+V_{s-1}^{(n+1)}+V_{s}^{(n)}\\
&  =-\rho_{1}V_{1}^{(n+s-2)}+2V_{2}^{(n+s-2)}+V_{3}^{(n+s-3)}+\dotsb
+V_{s-1}^{(n+1)}+V_{s}^{(n)}\\
&  =-\rho_{1}V_{1}^{(n+s-2)}-2\rho_{2}V_{1}^{(n+s-3)}+3V_{3}^{(n+s-3)}%
+\dotsb+V_{s-1}^{(n+1)}+V_{s}^{(n)}\\
&  =\dotsb=-\rho_{1}V_{1}^{(n+s-2)}-2\rho_{2}V_{1}^{(n+s-3)}-3\rho_{3}%
V_{1}^{(n+s-4)}-\dotsb-(s-1)\rho_{s-1}V_{1}^{(n)}+sV_{s}^{(n)}\\
&  =s\rho_{s}-\sum_{k=1}^{s-1}k\rho_{k}V_{1}^{(n+s-k-1)}=-\sum_{k=1}^{s}%
k\rho_{k}V_{1}^{(n+s-k-1)}.
\end{align*}
Again using (\ref{104}) and the inductive hypothesis we get
\begin{align*}
\sum_{k=1}^{s}V_{k}^{(n+s-k)}  &  =s\rho_{s}-\sum_{k=1}^{s-1}k\rho_{k}%
V_{1}^{(n+s-k-1)}=s\rho_{s}+\rho_{1}\sum_{k=1}^{s-1}k\rho_{k}V_{1}%
^{(n+s-k-2)}-\sum_{k=1}^{s-2}k\rho_{k}V_{2}^{(n+s-k-2)}\\
&  =s\rho_{s}+\rho_{1}Z_{s-1}-\sum_{k=1}^{s-2}k\rho_{k}V_{2}^{(n+s-k-2)}\\
&  =s\rho_{s}+\rho_{1}Z_{s-1}+\rho_{2}\sum_{k=1}^{s-2}k\rho_{k}V_{1}%
^{(n+s-k-3)}-\sum_{k=1}^{s-3}k\rho_{k}V_{3}^{(n+s-k-3)}\\
&  =s\rho_{s}+\rho_{1}Z_{s-1}+\rho_{2}Z_{s-2}-\sum_{k=1}^{s-3}k\rho_{k}%
V_{3}^{(n+s-k-3)}\\
&  =\dotsb=s\rho_{s}+\rho_{1}Z_{s-1}+\rho_{2}Z_{s-2}+\dotsb+\rho_{s-2}%
Z_{2}-\sum_{k=1}^{1}k\rho_{k}V_{s-1}^{(n-k+1)}\\
&  =s\rho_{s}+\rho_{1}Z_{s-1}+\rho_{2}Z_{s-2}+\dotsb+\rho_{s-2}Z_{2}%
+\rho_{s-1}Z_{1}=-Z_{s},
\end{align*}
where the last equality follows from the Newton's identity relating power sums
and elementary symmetric polynomials (see for example \cite{Macdonald}).

\section*{Acknowledgements}

Z. Doma\'{n}ski has been partially supported by the grant 04/43/DSPB/0106 from
the Polish Ministry of Science and Higher Education. M B\l aszak wishes to
express his gratitude for Department of Science, Link\"{o}ping, University,
Sweden, for their kind hospitality.

\end{document}